\documentclass{aa}  

\usepackage{graphicx}
\usepackage{lipsum}
\usepackage{txfonts}

\usepackage{float}
\usepackage{newfloat}
\usepackage{hyperref}
\hypersetup{colorlinks, linkcolor={blue}, citecolor={blue}, urlcolor={blue}}
\usepackage{afterpage}
\usepackage{amsmath}
\usepackage{subcaption}
\usepackage{siunitx}
\usepackage{booktabs}
\usepackage{multirow}
\usepackage{color}

\definecolor{raspberry}{rgb}{0.7,0.,0.3}
\definecolor{magenta}{rgb}{0.8,0,0.8}
\definecolor{purple}{rgb}{0.5,0,0.5}
\definecolor{gray}{rgb}{0.5,0.6,0.7}

\usepackage{silence}
\begin{document}

   \title{Large eROSITA X-ray sources as 2MRS galaxy groups}

   \author{A. Finoguenov \inst{1}
        \and
        R. Seppi \inst{2}
        \and
  D. Eckert\inst{2}
  \and
        H. Khalil \inst{1}
        \and
        J. Kosowski \inst{1}
          \and
        E. Tempel \inst{3,4}
        \and 
        F. Gastaldello \inst{5} 
        \and 
        L. Lovisari \inst{5,6}
          }

   \institute{Department of Physics,  University of Helsinki, Gustaf Hällströmin katu 2A, Helsinki, FI-00014, Finland 
         \and
       Department of Astronomy, University of Geneva, Ch. d’Ecogia 16, CH-1290 Versoix, Switzerland      
         \and
   Tartu Observatory, University of Tartu, Observatooriumi 1, 61602 T\~oravere, Estonia
\and
Estonian Academy of Sciences, Kohtu 6, 10130 Tallinn, Estonia
             \and
INAF - IASF Milano, Via E. Bassini 15, 20133 Milan, Italy
\and
Center for Astrophysics | Harvard \& Smithsonian, 60 Garden Street, Cambridge, MA 02138, USA
             }

   \date{Received June 21, 2025}

% \abstract{}{}{}{}{} 
% 5 {} token are mandatory

\abstract
 {Understanding baryonic physics at the scale of galaxy groups is essential for high-precision cosmological studies of the large-scale structure. Most models predict that the intergalactic medium in groups is hot and extends well beyond the virial radius, emitting X-rays on scales of up to several degrees in nearby systems. Early X-ray searches using ROSAT showed significant promise in recovering and identifying this emission.}
   {We aim to exploit the large area coverage, good sensitivity, and low instrumental background of eROSITA to detect the faint surface brightness emission of galaxy groups from the Two Micron All Sky Survey Redshift Survey (2MRS).}
   {Using the data from eROSITA-DE Data Release 1, including images, exposure maps, and local background maps, we performed a wavelet decomposition of image mosaics in the 0.6--2.3~keV band at angular scales of ${1\over 8}^\prime-16^\prime$. We adopted $8^\prime-16^\prime$ scales for source detection and $2^\prime-4^\prime$ scales to improve catalog purity. A novel identification method based on the ranked partial Hausdorff distance fully exploits the X-ray image and group membership information. Random catalogs were used to control match purity, and the identification threshold was chosen to maximize the catalog size at a fixed purity.}
   {We present a catalog of 619 X-ray galaxy groups with 80\% purity, and define subsamples with 90\% and 97\% purity. Bright sources closely match the AXES-2MRS catalog {(which is based on ROSAT All Sky Survey data analysis on spatial scales of $12^\prime-24^\prime$)}. The X-ray luminosity function of our groups agrees with previous studies down to $5\times10^{41}$~erg~s$^{-1}$. Using dynamical mass estimates, we find that the X-ray counterpart completeness for groups with $\ge 4$ members exceeds 60\% for masses $>2\times10^{13}~M_\odot$. We modeled the 2MRS group catalog and justify the inclusion of two-member groups in the identification.}
   {This study demonstrates that large X-ray sources on spatial scales relevant for cosmological studies of baryonic distributions can be reliably detected and identified using nearby galaxy group catalogs.}
   

\maketitle
\nolinenumbers
%
%-------------------------------------------------------------------

\section{Introduction}

{
The local Universe as mapped by the Two Micron All Sky Survey Redshift Survey \citep[2MRS;][]{2012ApJS..199...26H} provides a uniquely complete laboratory for studying group-scale environments, one that has underpinned a wide array of investigations — from constrained reconstructions of the cosmic velocity field \citep{2017MNRAS.469.2859S, 2018MNRAS.476.4362S} and neutral gas censuses \citep{2013AJ....146..124H} to analyses of galaxy properties in poor systems and the intragroup medium of fossil structures \citep{2009ApJ...696.1441T, 2014MNRAS.444..651M}. Crucially, the all-sky coverage of the 2MRS catalog, which extends into the Galactic disk, allows us to improve upon local dynamic studies that often mask or avoid this region. Over the past few decades, the 2MRS data have spawned a succession of group catalogs, beginning with the classical friends-of-friends construction of \citet{2007ApJ...655..790C} and numerous subsequent refinements \citep{2015AJ....149...54T, 2016ApJ...832...39L, 2016A&A...596A..14S, Tempel2017, 2017ApJ...843...16K, 2017MNRAS.470.2982L}.
On the theoretical side of halo formation, the standard hierarchical growth picture predicts self-similar behavior across the entire mass spectrum: purely gravitational simulations yield nearly universal density profiles for both dark matter and the intracluster medium when radial coordinates are normalized by the virial radius \citep{1995MNRAS.275..720N}. A direct corollary of this universality is the bremsstrahlung-dominated expectation 
for the X-ray luminosity–temperature scaling. In reality, however, observed clusters adhere to a significantly steeper 
 relation than this simple power law \citep{1997MNRAS.292..419W,1998ApJ...504...27M}, and the deviation becomes even more pronounced in the group regime \citep{2000MNRAS.319..933H}. \citet{1999Natur.397..135P} demonstrated that this steepening originates from a systematic suppression of central gas densities in poorer systems than in richer clusters. Framed in terms of the specific entropy of the intragroup medium, the breakdown of self-similarity translates into a well-established "entropy floor" -- an excess above gravitational predictions that persists outside the dense cooling cores, where radiative losses could otherwise erase it \citep{1999Natur.397..135P, 2000MNRAS.315..689L}. Explaining this floor has driven extensive theoretical work, with proposed mechanisms falling into three broad categories: early pre-heating of the gas before it collapses into group-sized halos \citep{1991ApJ...383..104K, 1991ApJ...383...95E, 2001ApJ...546...63T, 2003A&A...398L..35F}; in situ energy injection from supernovae or active galactic nucleus activity \citep{1997MNRAS.288..355B, 2000ApJ...532...17L, 2001Natur.414..425V, 2002MNRAS.333..145N}; and the selective extraction of low-entropy gas through radiative cooling, which effectively mimics a heating signature by biasing the remaining population toward a higher entropy \citep{1997MNRAS.289..955K, 2000ApJ...544L...1B, 2000MNRAS.317.1029P, 2001ApJ...552L..27M, 2002ApJ...569..112W, 2002ApJ...579...23D}. Although shock-driven "smooth accretion" during assembly remains a favored channel for setting this entropy baseline \citep{2003ApJ...593..272V, 2003Ap&SS.285..225B}, discriminating observationally among these scenarios demands a complete and unbiased census of nearby groups — precisely the regime where legacy surveys have proven inadequate.}
Several studies have also suggested a direct link between the baryonic content of galaxy groups and the shape of the matter power spectrum on spatial scales below 10~Mpc \citep{Debackere20}, and this information is required for the precision cosmology \citep{Semboloni11}. 
\begin{figure*}
%\centering
\sidecaption
   \includegraphics[width=12cm]{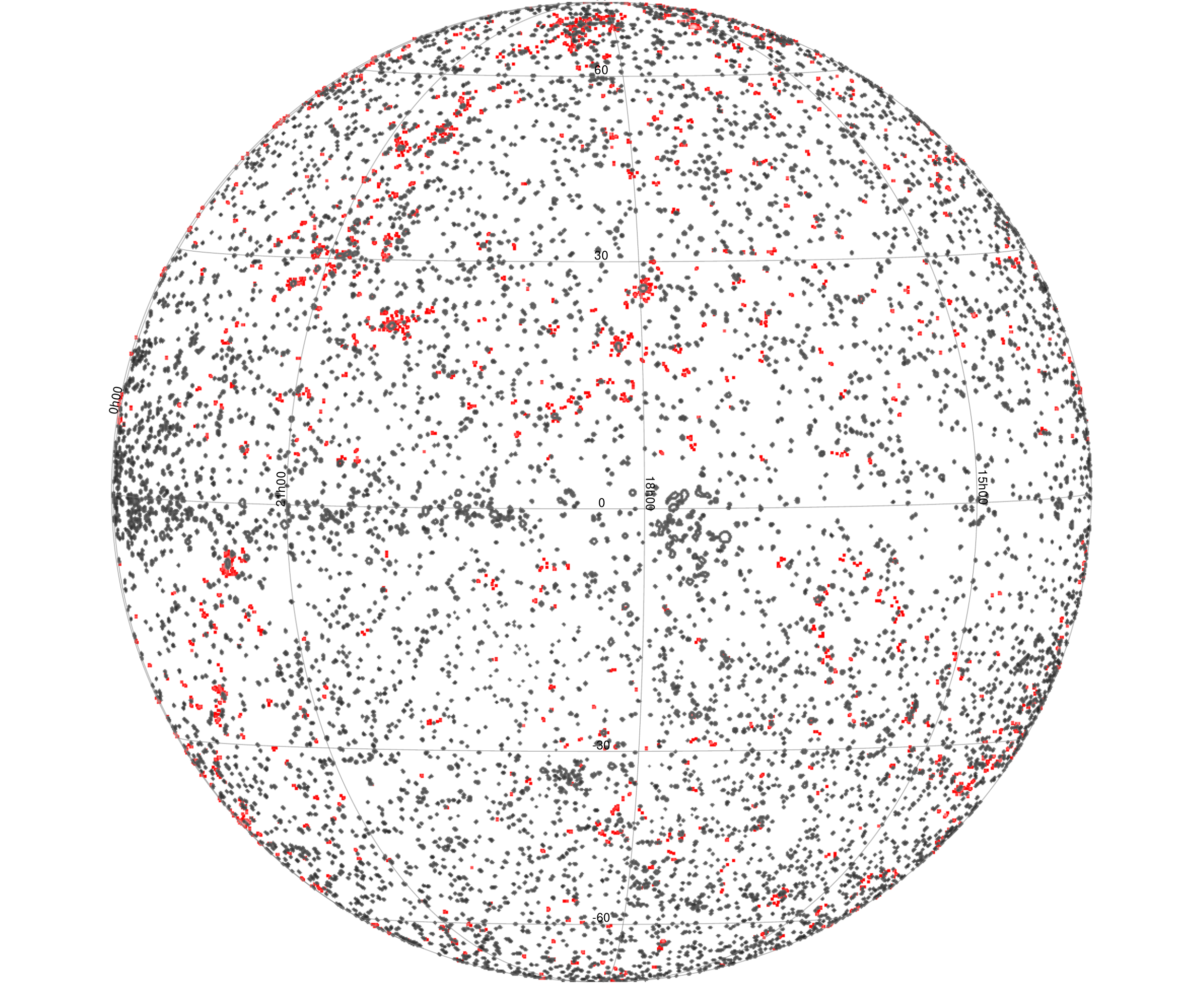}
     \caption{X-ray emission on angular scales of $8^\prime-16^\prime$ (gray contours) vs. 2MRS galaxy groups (red points show galaxy positions). The plot uses Galactic coordinates to show that most X-ray emission without a match to the 2MRS catalog is located in the Galactic plane. The 2MRS groups shown have at least three member galaxies.}
     \label{sgc}
\end{figure*}

Deep X-ray surveys have enabled significant advances in the understanding of galaxy groups, as they have discovered a large population of X-ray-emitting groups down to masses below $10^{13}~M_\odot$ and reaching redshifts above two for high-mass groups \citep{Gozaliasl2019}. However, the low-redshift population ($z<0.1$)  of galaxy groups is still not fully understood, even though it is a primary source of knowledge regarding the detailed properties of galaxy groups.
Previous catalogs of X-ray-selected local groups and clusters of galaxies have primarily been based on identifying sources encompassing an emitting zone of $2^\prime$ \citep[for details of ROSAT source detection, see][]{Voges99}. This has been shown to account for only a fraction of local galaxy groups, namely relaxed groups with luminous central objects \citep{mulchaey00}. A large population of sources is missing from these catalogs \citep{Xu18}, which has been confirmed by the dedicated study of galaxy group emission by \citet{Kaefer2019}. {The use of serendipitous and large-field survey catalogs based on \textit{XMM-Newton} only results in a handful of low-$z$ galaxy groups \citep{mehrtens, xxl18}. }

In a recent work we studied, using large spatial scales for detections, the X-ray emission from 2MRS \citep{2024A&A...690A.212K} and Sloan Digital Sky Survey (SDSS; \citealt{Damsted24, Kosowski25}) galaxy groups using ROSAT (ROentgen SATellite) All Sky Survey (RASS) data \citep{Truemper93}. We refer to the corresponding X-ray group catalogs as AXES-2MRS and AXES-SDSS. In this paper we continue the investigation of such sources by considering the spatially resolved X-ray emission on large spatial scales from our reanalysis of eROSITA (extended ROentgen Survey with an Imaging Telescope Array) data \citep[][]{2021A&A...647A...1P}.

In Sect.~\ref{sectiondata} we present the construction and basic properties of the new X-ray source catalog, and we describe the 2MRS optical group catalog used for the identification. We present the statistics of our catalog in Sect.~\ref{results} and provide a model for the joined detection in Sect. \ref{s:modeling}. We summarize our results in Sect.~\ref{conclusion}. In this study, we adopt a flat $\Lambda$ cold dark matter cosmology with the parameters $H_{0}=70$~km~s$^{-1}$~Mpc$^{-1}$, $\Omega_{\rm m}=0.3$, and $\Omega_{\Lambda}=1-\Omega_{\rm m}$. Unless otherwise stated, errors represent standard $1\sigma$ uncertainties (drawn at the 68\% confidence level).  
For radii, masses, and concentrations, the suffixes 200 and 500 correspond to the enclosed densities relative to the critical density of the Universe at the redshift of the group.
\section{Data} \label{sectiondata}
\subsection{eROSITA data selection}
The analysis utilizes the portion of the sky with Galactic longitude $l>179.6^\circ$, released as a part of eROSITA-DE Data Release 1 \citep[eRASS1 for short, although eRASS1 refers to the first full scan of the sky;][]{2024A&A...682A..34M}\footnote{\url{https://erosita.mpe.mpg.de/dr1}}. The data are organized in overlapping mosaics, each covering approximately 10 square degrees. For our study, we selected the band 022 (0.6--2.3~keV). We avoided the softer energy band (e.g., included in band 024, 0.2--2.3~keV) for two primary reasons: First, it suffers from a high level of foreground due to charge exchange and the Local Bubble \citep{2014Natur.512..171G, 2026Sci...392..285D, 2026arXiv260502998Y}, as well as possible confusion with large-scale filaments, which are reported in recent eROSITA analyses of soft-band emission \citep{2024A&A...691A.286D}. Second, it is more susceptible to contaminants, such as reported solar light leaks from two of the telescopes \citep{2021A&A...647A...1P}. A direct comparison of source detection in bands 022 and 024 confirmed that band 022 is superior for the spatial scales adopted in our analysis.
Given that most of the data comprise a single observational pass, screening based on background levels does not improve source detection. Therefore, we performed our analysis directly on the publicly released images, exposure and background maps.

\subsection{Wavelet analysis} 

The officially released eRASS1 background maps, which employ an in-field estimate smoothed on a scale of half a degree, impose limitations on our analysis. While necessary to mitigate edge effects in wavelet analysis, this smoothing limits the analysis to smaller scales, setting a practical limit for wavelet decomposition to $16^\prime$. Another feature of the standard background consists of over-prediction of the emission near bright sources, which we discuss in Appendix A. 

{To optimally recover the extended, low-surface-brightness emission characteristic of nearby systems, we tuned the spatial scales for the X-ray source detection to angular scales of  $8^\prime-16^\prime$, chosen to bracket the virial radii of groups at $z<0.1$. We used a wavelet decomposition of \citet{vikhlinin98}\footnote{\url {https://github.com/avikhlinin/wvdecomp}}, which provides a multi-scale detection, unaffected by the emission on scales different to the scale of interest. The choice of angular scales for our X-ray detection is designed
to cover the virial radius of groups at $0.01<z<0.1$, which corresponds to the angular scales in the  $2^\prime-16^\prime$ range. The detection threshold is set to four times the background noise level.}   
The eROSITA survey-averaged point spread function (PSF) has a half-energy width of $0.5^\prime$ \citep{2021A&A...647A...1P}. In extracting the source flux, we excised the area of sources detected on spatial scales of $0.5^\prime$ or less. {These are 664 000 unique source areas (counting blended sources within an area as one), of which 566 000 contain literally all significant unresolved sources present in the \cite{2024A&A...682A..34M} catalog}. This procedure also results in a removal of cool core emission, in a similar fashion to our analysis of deep X-ray fields, making our calibrations using weak lensing mass estimates \citep{Leauthaud2010}  directly applicable. {In \cite{2025ApJ...988..238P} we also provide direct weak lensing calibration of our eROSITA luminosity estimates and a comparison to ROSAT.} 

For the source identification, we followed the approach developed for identifying RASS sources using SDSS groups \citep{Kosowski25}. The method relies on the identification of groups using their outskirts. The definition of a uniform baryonic overdensity in the RASS analysis corresponds to an eROSITA count rate of $10^{-6}$~counts~s$^{-1}$~pixel$^{-1}$ in the 0.6--2.3~keV band with $4^{\prime\prime}$ pixels and extragalactic nH conditions. We stored the contours of X-ray emission from the wavelet decomposition on scales of $2^\prime-4^\prime$ and $8^\prime-16^\prime$. The contours were obtained using the {\it ds9} program through the {\it xpa} interface\footnote{https://ds9.si.edu/doc/ref/xpa.html}, applying a smoothing of two pixels.

The contour defines the flux extraction zone, which is obtained by converting the {\it ds9} contour file to a polynomial region. We assigned an ID to each contour composed of the ID of the field and the numbering of the contours within the field. Given the overlap between the sky tiles of eROSITA, we have duplicate sources, which were identified using the Hausdorff distance, as described in detail below; the contour with the largest signal-to-noise ratio was kept. In addition, incomplete contours, coming from seeing only part of a source in a tile, were removed entirely. These were identified by a large distance between the starting and ending points of the contour compared to a typical interval between contour points.

In total, we have 10,000 sources on scales of $8^\prime-16^\prime$ and 40,000 sources on scales of $2^\prime-4^\prime$. We could efficiently control point source contamination by computing the ratio of extended to point source flux to account for possible PSF contribution. Using the full catalog of sources, we measured the saturation level of the total to point source count rate ratio, identifying a contribution of compact sources as 0.5 of their flux encompassed within $0.5^\prime$ radius aperture, which is consistent with the eROSITA PSF. We subtracted this contribution from the scattered light from point sources and removed sources with residual flux below $2\sigma$. With this cleaning the number of unique large sources reduces to $\sim 6,300$ and small sources to $\sim 27,300$.   

Comparison with the public eRASS1 cluster catalog \citep{erass1_new} using  $1^\prime$ matching radius shows that among the 12,000 clusters identified using redMaPPer, 10,000 are detected by us on scales of $1^\prime-2^\prime$, 6,000 are detected on scales of $2^\prime-4^\prime$, while only 17 clusters are in common with our detection on scales of $8^\prime-16^\prime$. Since those spatial scales are not in the official detection pipeline, we discuss the results using much larger matching radii in Sect. \ref{s:modeling}. {Source selection in the public eROSITA catalogs requires a peak detection after a smoothing of the image with a $15^{\prime\prime}$ kernel, after that the source is checked on the extent \citep{2022A&A...661A...1B}. This procedure is suitable for distant galaxy clusters on the edge of the instrumental resolution, but does not cover the physical scales of the intragroup emission at low redshifts \citep{Kaefer2019}. Our demonstration of the large source overlap on small scales reveals a consistency in the results of the source detection when the same spatial scales are used.}

\subsection{Source identification methodology}

{For the present X-ray study, we adopted the catalog of \citet{tempel18}, which employs a Bayesian marked point process -- a framework that, while mathematically distinct from friends-of-friends, produces partitions that closely mirror the traditional linkage output in practice \citep{2016A&A...588A..14T}. We drew the parent photometric and spectroscopic sample from \citet{2012ApJS..199...26H}, selecting galaxies brighter than 11.75~mag in the $K_\mathrm{S}$ band. It is highly complete above the Galactic plane (Galactic latitudes $|b|>5^{\circ}$). To counteract the increasing sparsity at larger distances, a 300 Mpc distance cut was imposed, leaving 42,620 galaxies that are assembled into 7,755 groups (of which 1,933 contain at least three spectroscopic members). A persistent conceptual challenge in group-finding is whether a recovered structure reflects a genuine physical overdensity or an algorithmic artifact imposed upon the data. The Bayesian approach adopted by \cite{tempel18} is particularly attractive precisely because it models the spatial regions that define group membership, rather than merely linking discrete point detections \citep{2005A&A...434..423S, 2010A&A...510A..38S}, thereby offering a richer morphological and statistical characterization of the clustered pattern. }

Figure~\ref{sgc} displays the X-ray contours describing the emission on scales of $8^\prime-16^\prime$ overlaid on the member galaxies of groups having  at least three spectroscopic member galaxies, all projected in Galactic coordinates. This projection clearly highlights the association of foreground diffuse emission with the Galactic plane, while the prominent eROSITA bubble is also readily visible near the left edge, near zero galactic longitude.

Our source identification follows the approach developed for RASS-SDSS groups \citep{Kosowski25}, which matches the 2D shapes of X-ray emission to optical galaxy groups using a modified Hausdorff distance (MHD) metric. This method emphasizes the identification of groups via their outskirts. This matching was performed separately for the $2^\prime-4^\prime$ and $8^\prime-16^\prime$ wavelet scales.
In \cite{Kosowski25} we introduced 

\begin{equation}
    d_{\mathrm{mod}}(X,Y)=\max\left[\underset{x\in X}{\mathrm{P50}} \left[\min_{y\in Y}\lVert x-y\rVert\right]\right] \ .
    \label{eq:mhd_dir}
\end{equation}

\noindent for two finite point sets, $X=\{x_1,x_2,...x_i\}$ and $Y=\{y_1,y_2,...y_i\}$, where P50 denotes the lower 50\% of the distribution, and using full distribution (P100) in this notation corresponds to the original definition of Hausdorff distance. {We traced our approach to the ranked partial Hausdorff distance, as formalized in \cite{Huttenlocher1993}. We adopt MHD notation henceforth.}

In addition to the fractional condition, we required that the size of the matched sample be larger than some threshold value and accounted for the probabilistic membership associated with the vector $X$ galaxies $\{p_1,p_2,...p_i\}$. We tested the thresholds of 2, 3, and 5, finally selecting 2, which is lower than what we assumed in AXES-2MRS catalog construction \citep[three member groups were used there;][]{2024A&A...690A.212K}, but this is required by the larger depth of eRASS1 compared to RASS  data, as illustrated by our modeling of 2MRS groups in Sect. \ref{s:modeling}.

\begin{equation}
P50: {\underset{x\in D}{\sum p_i} } \ge max(2,  0.5  {\underset{x\in X}{\sum p_i}} )
,\end{equation}
%\begin{equation}
%P50: {\underset{x\in D}{\sum p_i} } \le 0.5  {\underset{x\in X}{\sum p_i}}
%\end{equation}

\noindent where $D$ is a subset of X that satisfies
\begin{equation}
\min_{y\in Y}\lVert x-y\rVert \le d_{\mathrm{mod}}(X,Y).
\end{equation}

\begin{figure}[hbt!]
\centering
   \includegraphics[width=\columnwidth]{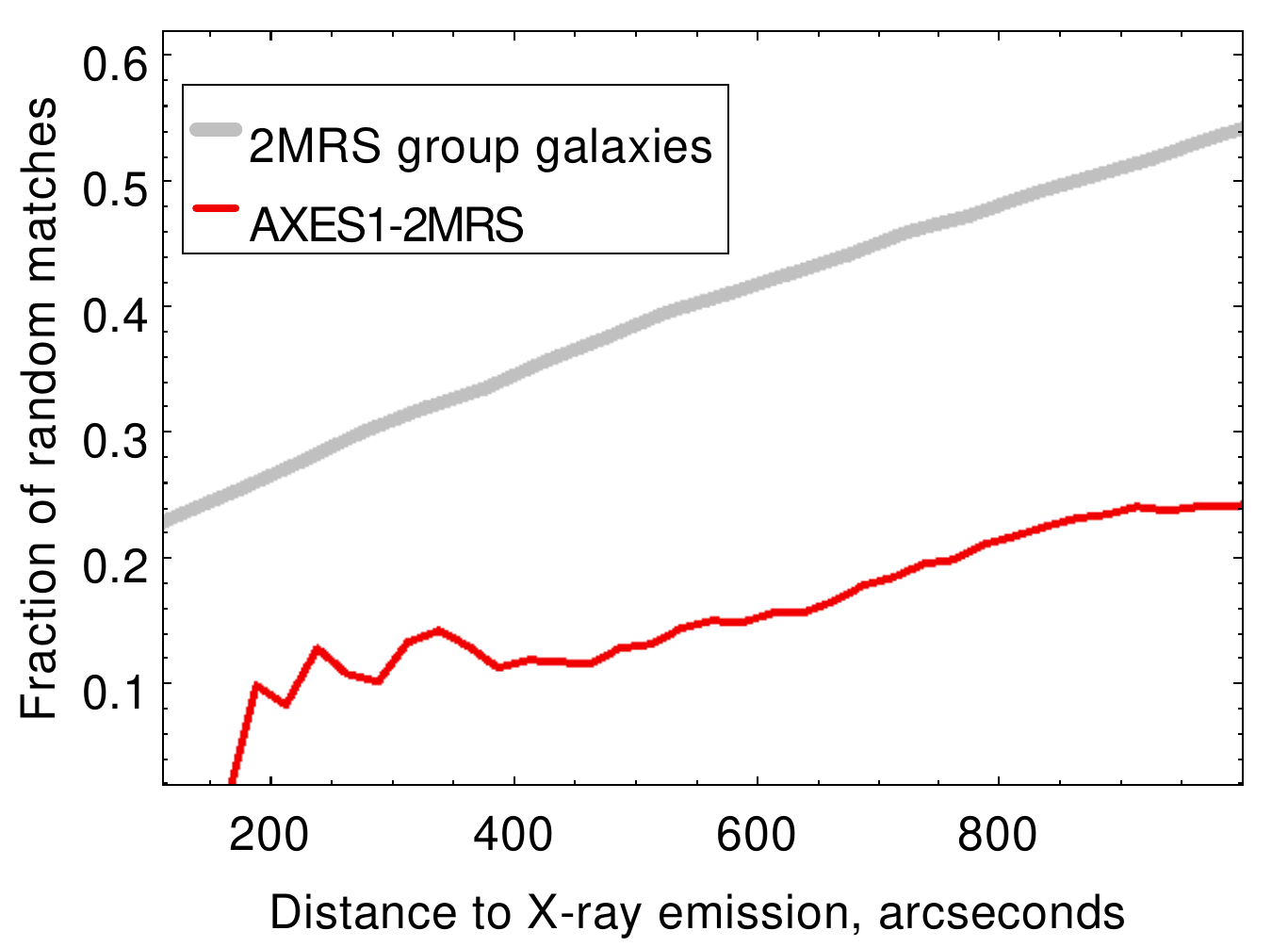}
     \caption{Cumulative fraction of random matches as a function of a distance to the X-ray emission. The gray curve shows the galaxies, and the red curve shows the ranked partial Hausdorff distances for eRASS1-2MRS galaxy groups.}
     \label{g2x}
\end{figure}

We searched for a minimal value of $d_{\mathrm{mod}}(X,Y)$ that satisfies this criterion. The total number of {2MRS} groups that can potentially be matched using this criterion is 2850 in the eRASS1 sky since a number of two member groups have summed membership probabilities below 2.   
For practical convergence, we excluded galaxies with membership probabilities below 1\%. We applied the criteria of having two (or more) galaxies to the X-ray emission map. The fractional criterium was applied to the X-ray contour as a posterior cleaning step, as it is not driving the overall contamination but is important in reduction of random matches for rich systems. Using the absolute number of galaxies in the match closely mimics the standard practice of X-ray identification \citep[see, e.g.,][]{Kirkpatrick2021}, which is what we would like to implement and test. These tests showed that large groups have a high probability of producing such an identification by chance, and our proposed solution is to complement this identification using the Hausdorff criteria. On the other hand, without imposing a requirement on the absolute number of galaxies, some identification could be produced by a single galaxy, which we will consider in a separate work and has been been studied by \cite{2025A&A...697A.173C}.
To evaluate the chance associations, we created a random group catalog by shifting the right ascension of all real groups by 10 degrees (modulo 360). {Figure~\ref{g2x} compares the fraction of randomly matched galaxies to the fraction of randomly matched groups as a function of their distances to X-ray emissions (for groups we followed the definition of the distance introduced below).} With our goal to reach the final level of contamination below 30\%, the symmetrical contribution of the probability of each term yields a targeted value of contamination of 55\%, while we sampled values from 30\% to 98\%, {limiting the maximal distance to X-ray emission to $1200^{\prime\prime}$. We built our matching scripts on STILTS \citep{2005ASPC..347...29T}, which is why we implemented a consequent refinement scheme.}

For the next step of matching X-ray contours to galaxy groups, we used only the member galaxies selected to be close to X-ray emission in the previous step.
Our matching criteria are based on two parameters: (1) the required fraction of contour points that must be matched and (2) the maximum allowed distance for a point to be considered a match. 
In \cite{Kosowski25} we used a fixed fraction of the contour points in the match (50\%), which leads to the following criterion:

\begin{equation}
    d_{\mathrm{mod}}(Y,X)=\max\left[ \underset{y\in Y}{\mathrm{P50}} \left[\min_{x\in D}\lVert x-y\rVert\right]\right]  \ ,
    \label{eq:mhd_dir2}
\end{equation}

\noindent where a subset $D$ is fixed at a previous step. In our procedure we tested percentiles P10, P30, P50, P65, P80, P90, P95, and P100.
The procedure samples all values of both $d_{\mathrm{mod}}(Y,X)$ and $d_{\mathrm{mod}}(X,Y)$, producing catalogs of real and random groups that satisfy the selection criterion, and finds the maximal distances that yield the required ratio between true and random catalogs, associated with a purity level (set in the range 70--95\%).

\begin{figure}[hbt!]
\centering
   \includegraphics[width=\columnwidth]{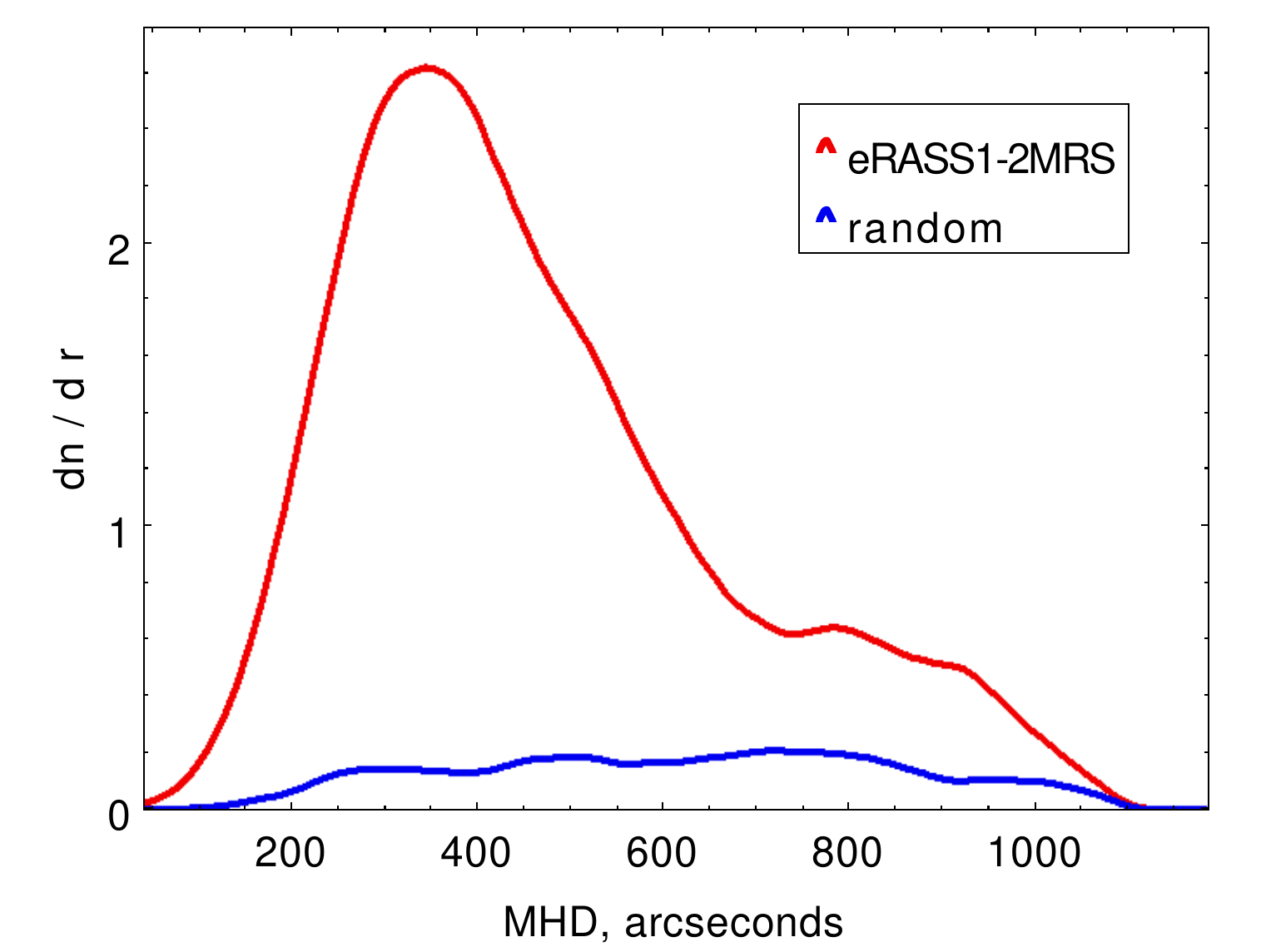}
     \caption{Distribution of the ranked partial Hausdorff distances for galaxy groups. The red curve shows the eRASS1-2MRS catalog, the blue curve our estimate for the random matches.}
     \label{mhd}
\end{figure}

To ensure the match reflects the contour's shape, the maximum distance must be a small fraction of the contour's overall size. If this distance were too large (e.g., on the order of the contour's full size), matching a single point would effectively validate the entire contour, which immediately boosts the random association. 
We also find that requiring a larger fraction of matched contour points reduces the likelihood of chance associations. 
We did not require a contour to match a single optical group; instead, we required that the optical counterparts collectively describe the majority of the X-ray source.  The values of the ranked partial Hausdorff distance describing these matches are in the range $15^\prime-20^\prime$ for scales $8^\prime-16^\prime$, but for small-scale detections ($2^\prime-4^\prime$), MHD values are $3^\prime-6^\prime$. The values of the distances depend critically on the size of the optical catalog; for example, using the redMaPPer membership catalog down to richness 3 sources, we used distances that were a factor of 4 smaller due to the large number of sources \citep{2026arXiv260702706D}. So, large distances employed in this analysis are unique to the 2MRS group catalog and offer certain advantages in terms of completeness of the matching procedure, as discussed in \cite{Kosowski25}. For comparison, the median intergalactic separation of the 2MRS group catalog in the $0.015<z<0.04$ range is $4^\prime$, with 90\% of separations in the $1^\prime-10^\prime$ range. Such separations imply that we have a good match between the group sizes and the spatial scales of X-ray emission. {In Fig. \ref{mhd} we show the distribution of the group to X-ray contour distances for the full catalog reported in this paper. The purity constraints still allowed us to fully capture the main peak of the distribution, which is the root to a high fractional match of this sample, discussed below. Use of MHD matching allowed us to reduce the fraction of random matches compared to galaxies, as illustrated in Fig.~\ref{g2x}. To capture the main peak of associations, we needed to accept MHD distances of up to $600^{\prime\prime}$, which corresponds to purity level in the 80-90\% range. The MHD distribution for the small-scale catalog is similar, but the MHD values had to be cut by a factor of 3  due to contamination and; as a result, it does not capture all matches.}

We find that at purity levels below 80\% there is a problem of multiple associations between X-ray and optical sources, complicating statistical evaluations of completeness. We therefore selected the 80\% purity calculations to illustrate the catalog performance. For the detailed study of scaling relations using this catalog, we used the 90\% pure catalog, and in the released catalog we introduce a column indicating which groups belong to the 90\% purity subset, while the results of these studies are presented elsewhere \citep{Khalil26} and are the subject of an ongoing large \textit{XMM-Newton} program (PIs: Lovisari, L. \& Gastaldello, F.).

\begin{figure}[hbt!]
\centering
   \includegraphics[width=9cm]{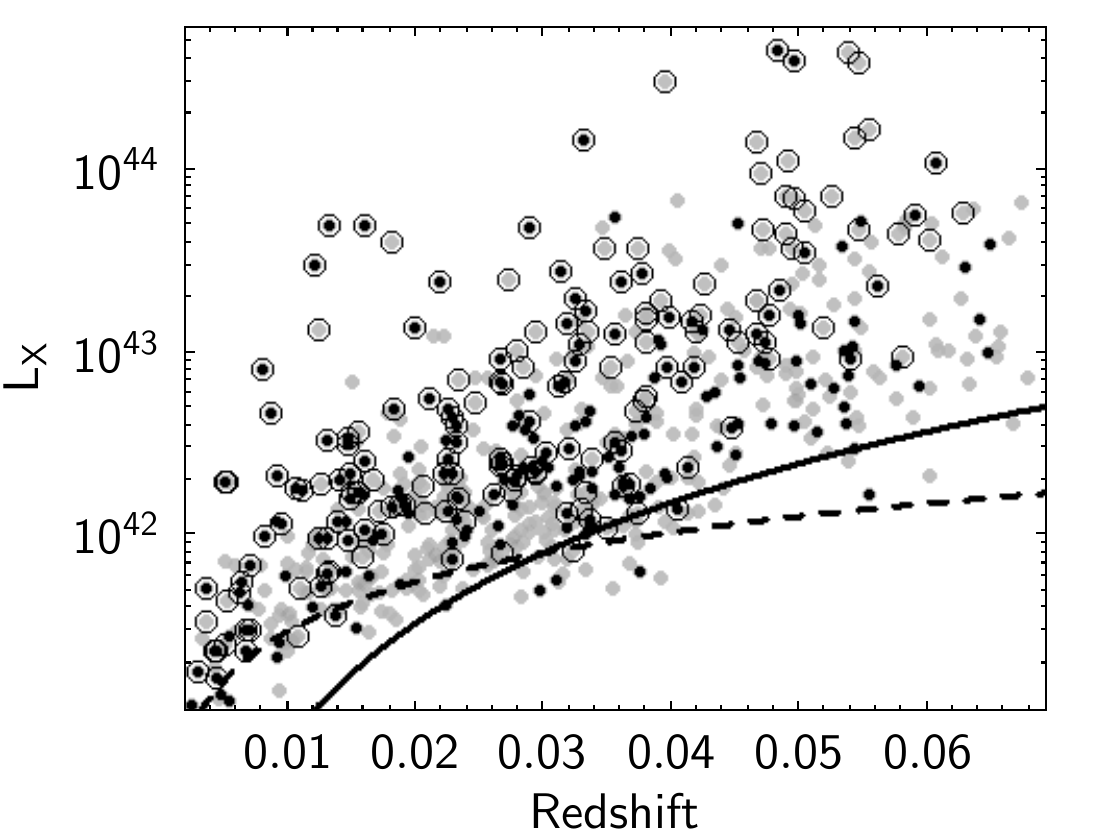}
     \caption{$L_{\rm X}$--$z$ plane of the eRASS1-2MRS catalog. Gray points show the catalog obtained on scales of $8^\prime-16^\prime$ and black points indicate detections on scales of $2^\prime-4^\prime$. The computation of X-ray luminosity uses the corresponding aperture flux. Black circles show matches to the AXES-2MRS catalog, which used even larger scales and RASS data. Black curves show analytic approximations to the sensitivity limit of the survey (dashed: $z<0.03$, solid: $z>0.03$).}
     \label{zlx}
\end{figure}

We then computed the X-ray flux and luminosity for each source. This calculation incorporates: $K$-corrections for the observed 0.6--2.3~keV and the rest-frame 0.1--2.4~keV bands; the exact hydrogen column density ($n_{\rm H}$) toward each source; an iterative temperature estimate derived from the luminosity--temperature relation from \cite{1998ApJ...504...27M}; and a correction for missing flux based on the source's effective radius, following \cite{finoguenov07}. At $z>0.03$ our flux extraction apertures cover the estimated $R_{500}$, while at $z<0.015$ we extracted the flux below $0.4 R_{500}$, which is not ideal for detecting group outskirts. 

\begin{figure}[hbt!]
\centering
   \includegraphics[width=\columnwidth]{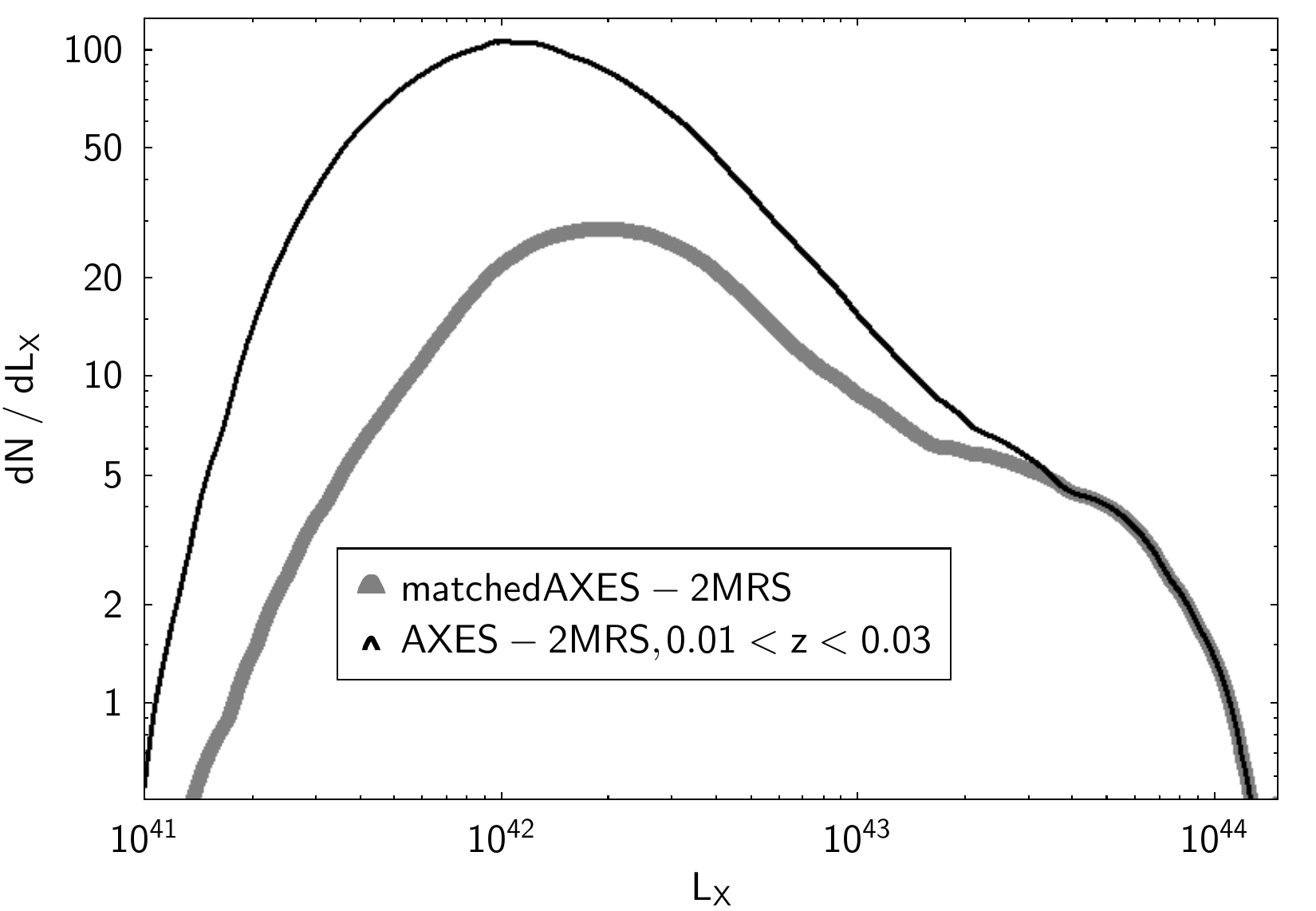}
     \caption{{Comparison of the luminosity distributions for eRASS1-2MRS (black curve) and what is matched to the AXES-2MRS sample (gray curve) in the redshift range $0.01<z<0.03$. We demonstrate that the eRASS1 catalog is deeper by a factor of 2 based on the position of the peak of the distribution.}}
     \label{lxdist}
\end{figure}

\begin{figure}[hbt!]
\centering
   \includegraphics[width=\columnwidth]{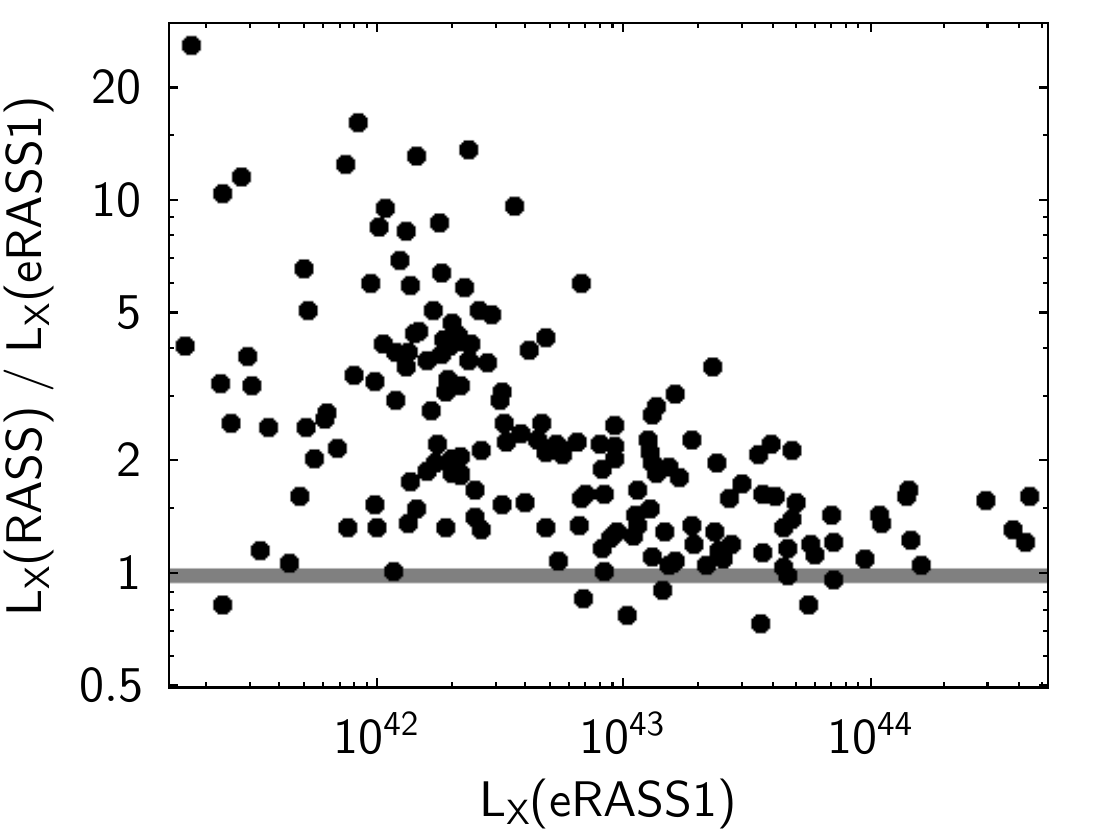}
     \caption{Comparison of $Lx$ measurements from eRASS1 and RASS for the sources in common. We see that at $L_{\rm X}<5\times10^{42}$ erg s$^{-1}$ eRASS1 provides a much cleaner $Lx$ measurement due to the implemented removal of resolved sources (RASS flux calculation is based on the full aperture flux) and smaller flux apertures \cite[RASS apertures often include the flux outside of $R_{500}$, as discussed in detail in][while eRASS1 background maps include this emission and therefore an excess emission is reported]{Khalil26}.}
     \label{lxcmp}
\end{figure}

In Fig.~\ref{zlx} we present our sample of 619 eROSITA identifications of 2MRS groups, comparing them to our previous catalog of ROSAT identifications from AXES-2MRS \citep{2024A&A...690A.212K}. We matched the sources based on their 2MRS group ID. Our new catalog extends the X-ray characterization of groups toward lower X-ray fluxes, and our new identification method allows us to recover a few additional high-luminosity systems, as previously demonstrated by \cite{Kosowski25}. eROSITA data are deeper and deliver a completeness that is a factor of 2 better, as illustrated in Fig.~\ref{lxdist}, and, most importantly, provides a much cleaner measurement of the source flux contained within $R_{500}$ (Fig.~\ref{lxcmp}), which is important for planning the follow-up. This effect has been modeled by \cite{Seppi2025A&A...699A.206S} and  verified by \cite{Khalil26}, who find that most of the ROSAT excess flux comes from radii outside of $R_{500}$. We also show the groups with detection on small ($2^\prime-4^\prime$) scales. Note that determination of the total flux of a group using small scales is subject to larger extrapolations, which affect the positions of groups on the plot. In performing the source identification, we realized that a large number of identifications is produced by two-member groups. We discuss this in more detail in the modeling of 2MRS sensitivity (Sect. \ref{s:modeling}).

The functional form of the completeness in terms of intrinsic 0.1--2.4~keV luminosity is a broken power law with a knee at $z=0.03$. The part for $0.005<z<0.03$ is $8\times 10^{41} \times (z/0.03)^{0.9}$~erg ~ s$^{-1}$ and for $0.03<z<0.07$ it is $8\times 10^{41} \times (z/0.03)^{2.2}$~erg ~ s $^{-1}$, calculated using a single flux limit of $3\times 10^{-13}$~erg ~ s $^{-1}$~cm$^{-2}$ and a flux aperture of $8^\prime$. At $z<0.03$, corrections for the missed source flux are dominant in determining the scaling with redshift, whereas at $z>0.03$ the $K$-correction is dominant. Detections below the plotted sensitivity curve come from the deeper part of the survey near the southern ecliptic pole. Small scales are not advantageous for source detection in our redshift and flux range, as the surface brightness profiles of galaxy groups are shallow in the spatial zone covered by the detection scales \citep{Khalil26}.

In comparing the catalogs using small ($2^\prime-4^\prime$) and large ($8^\prime-16^\prime$) spatial scales, we noticed that there is a large correspondence in the real detections, but not in the random identifications. This makes the sample of groups identified with both X-ray catalogs very pure, about 97\%, while a combined catalog using a detection at one of the scales would have a very low purity (60--70\%). {In the following we refer to the large-scale detections as the "main catalog." The subset that also has small-scale counterparts is referred to as the "intersection sample."} In the released catalog, we mark systems that are also detected using the small scales. Further exploration of the small-scale catalog will be presented elsewhere \citep[e.g., for the source identifications at higher redshifts, see][]{2026arXiv260702706D}. If several counterparts for a single X-ray source are available, we mark the identification as contaminated. A high purity level of association between the optical group and the X-ray emission indicates a correct assignment of the emission. Its calculation takes into account the fraction of sources that remain unidentified, which boosts the chance probability of an association. Our cleaning of the source catalog to remove the sources that can be explained as PSF wings of bright point sources led to a factor of 1.5 reduction in the contamination. Contamination coming from distant groups and clusters is traced by the small-scale emission, which is not used in the detection, and statistically forms only 10\% of the active galactic nucleus contamination. As we noted, the detection and association on both small and large scales with the same system led to the cleanest catalog. The detection on all scales used in this work is not confusion-limited. Formally speaking, only the emission features identified with the group shall be included in the flux estimates. The purity of 97\% implies that only in 3\% of cases (mostly lowest flux in the catalog, based on the random catalog), with deeper data, the identification of the source might change, as we have seen in X-GAP follow-up. In the final catalog we only release the properties of groups obtained using large ($8^\prime-16^\prime$) scales.

\begin{figure}[hbt!]
\centering
   \includegraphics[width=\columnwidth]{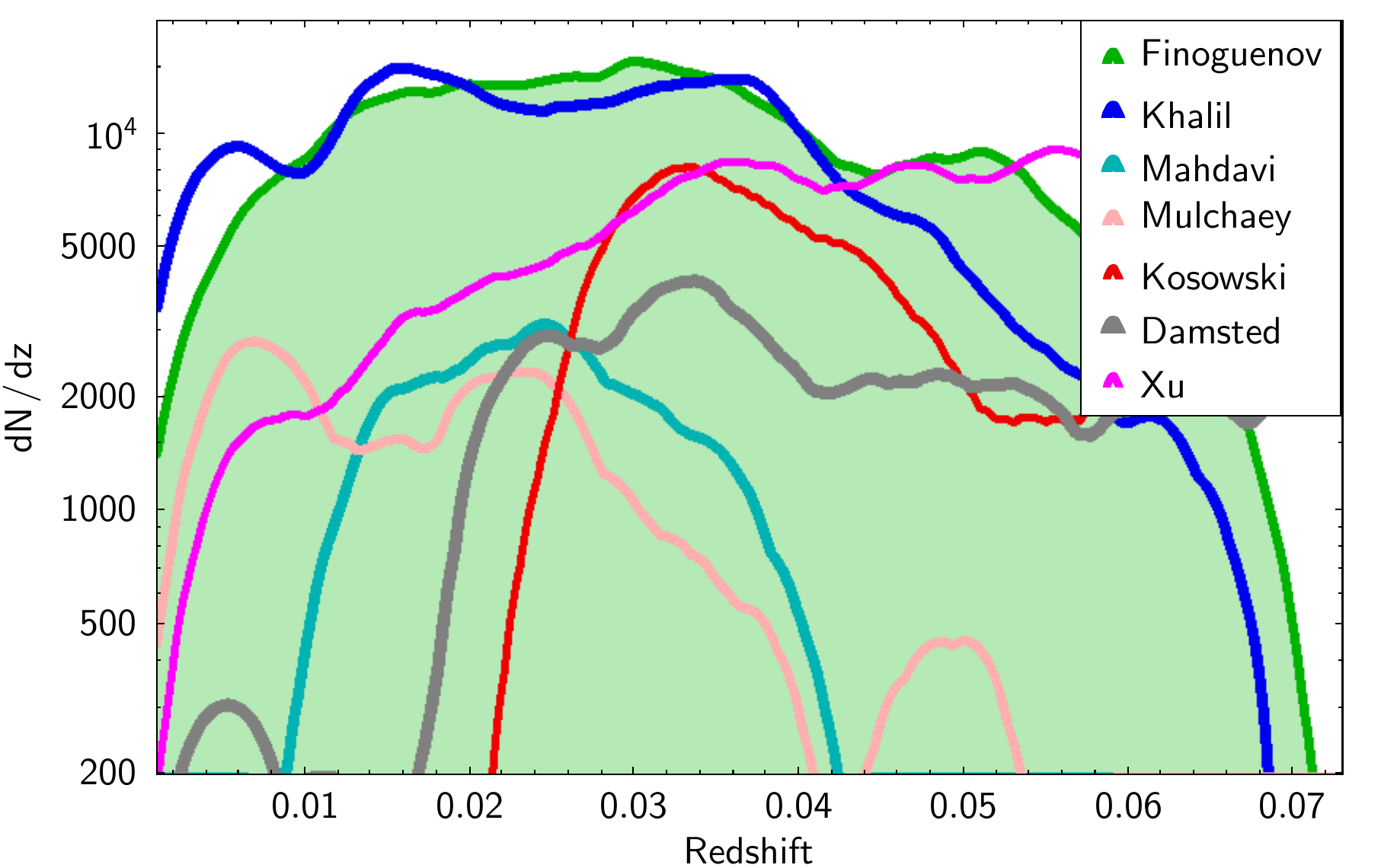}
     \caption{{Redshift distribution, $dN \over dz$, of largest low-$z$ group catalogs with extended X-ray emission. The eRASS1-2MRS catalog is shown in green (line and shaded area), and the AXES-2MRS catalog in blue. For comparison we show the previously known groups with extended X-ray emission from \citet[][pink curve]{2003ApJS..145...39M} and \citet[][cyan curve]{2000ApJ...534..114M}. Other catalogs of extended X-ray sources associated with groups and clusters of galaxies coming from reanalyses of RASS data are added for completeness: SDSS-based catalogs of \citet[][gray curve]{Damsted24} and \citet[][the intersection catalog is shown in red]{Kosowski25}, which are comparable at $z>0.03$, and a catalog from \citet[][pink curve]{2022A&A...658A..59X}, who used public redshift data to select systems containing three spectroscopic redshifts inside the X-ray-emitting zone.} }
     \label{gsamples}
\end{figure}

\begin{figure}[hbt!]
\centering
   \includegraphics[width=\columnwidth]{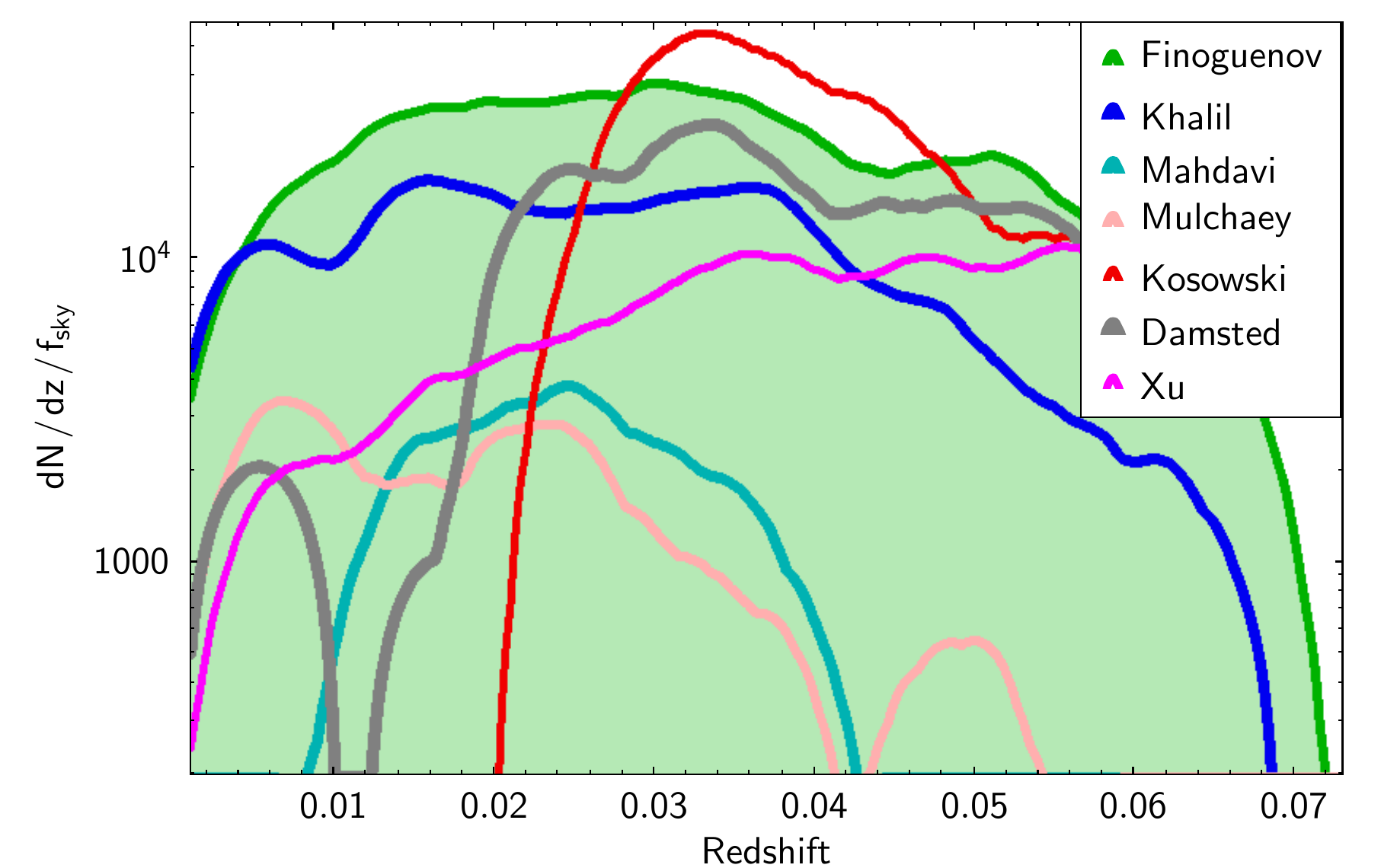}
     \caption{{Redshift distribution of galaxy groups normalized by the fraction of the sky occupied by each survey. Labeling is the same as in Fig. \ref{gsamples}. SDSS samples perform better above redshift 0.03 even with RASS data. We used a 6000 square degree area from the SDSS spectroscopic survey and sky fractions of 0.82 for other extragalactic surveys (0.41 for eRASS1-2MRS).} }
     \label{gsamples2}
\end{figure}

{
In Figs. \ref{gsamples} and \ref{gsamples2} we compare eRASS1-2MRS and AXES-2MRS catalogs with other X-ray group catalogs available in the literature. Our use of large scales of X-ray emission in combination with 2MRS groups yields the largest catalogs below redshifts of 0.05. Compared to the well-studied catalogs of X-ray groups of \cite{2003ApJS..145...39M} and \cite{2000ApJ...534..114M} (by \citealt{2005ApJ...622..187M, 2006ApJ...646..143F, 2007MNRAS.374..737F, clogs, Lovisari15, Lovisari21}), the new samples  (which include a similar analysis of RASS data by \citealt{2022A&A...658A..59X}, \citealt{Damsted24}, and \citealt{,Kosowski25}) are a factor of 5 larger even at the same redshift, which opens a new territory, with the first detailed insights into the properties of the intragroup media available from the \textit{XMM-Newton} follow-up studies of \cite{2025A&A...700A.220S} and the X-GAP survey \citep{2026A&A...710A.153S}.}

Summarizing the results of the identification, we have extended our new developments on the source identification \citep{Kosowski25} to eROSITA data. Compared to the original paper of \cite{Kosowski25}, we have extended the application of the method to low number of galaxy group members. We see that a sky density of sources at $z>0.03$ is even higher in AXES-SDSS, which means that our survey is limited by the depths of the spectroscopic data. We used this to our advantage as the deeper data allowed us to fully trace the variety of X-ray properties of optically selected groups. Compared to AXES-2MRS \citep{2024A&A...690A.212K}, we detected a larger density of sources with the eRASS1-2MRS catalog, except for at $z<0.0006$, where aperture flux losses compensate for the higher sensitivity of eROSITA. The great deal of consistency between RASS and the two scales of eRASS1 detections is further proof of the high purity of the catalog presented. This catalog offers a factor of five boost in the sample size of local X-ray galaxy groups, which we started exploring with the X-GAP survey \citep{xgap}, and provides much cleaner X-ray measurements compared to RASS data.

\section{Properties of the eRASS1-2MRS catalog}\label{results}

\begin{figure}[hbt!]
\centering
   \includegraphics[width=\columnwidth,trim={0 6cm 0 6cm},clip]{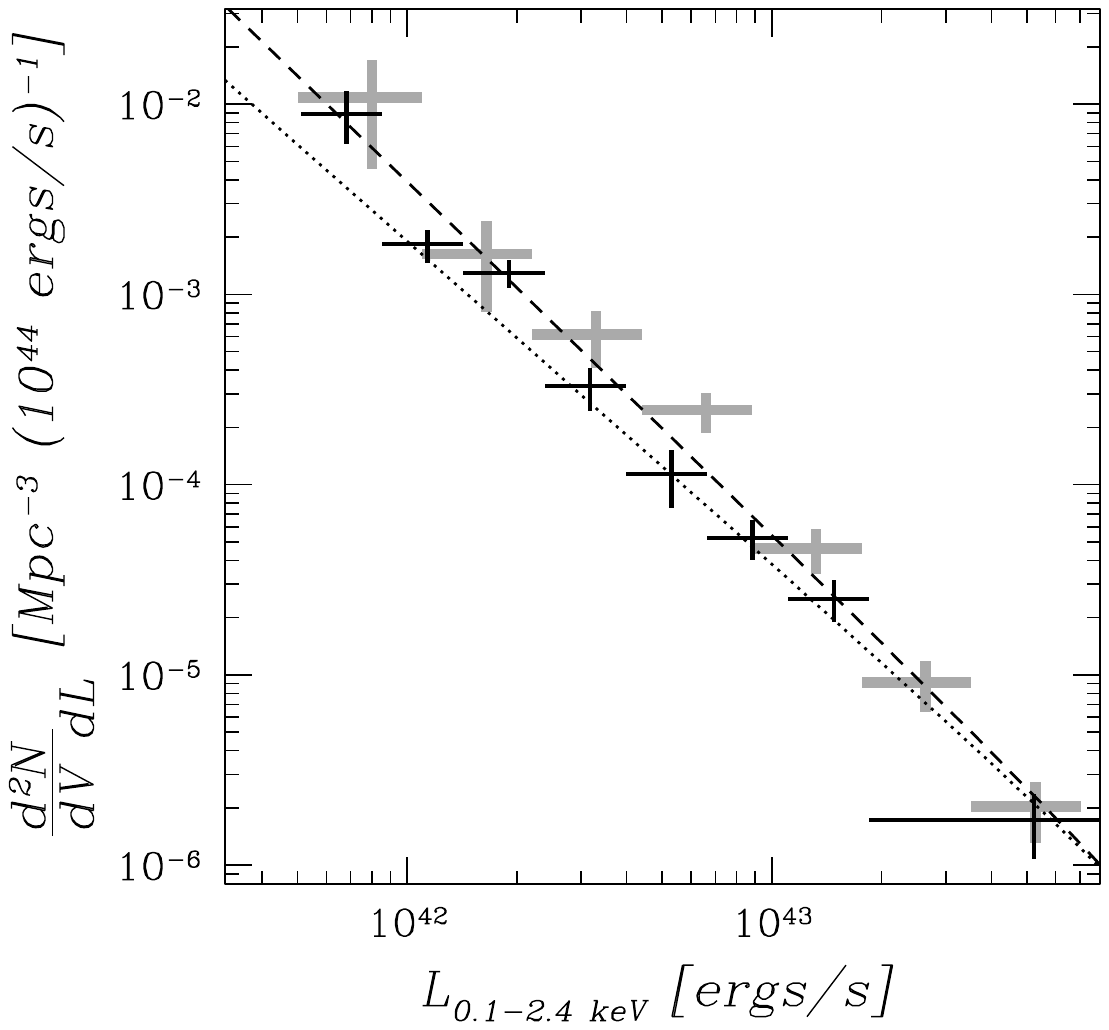}
     \caption{XLF. Estimates based on the main eRASS1-2MRS catalog are shown as black crosses. Gray crosses show XLF measurements in COSMOS \citep{finoguenov07}. The dotted line shows the luminosity function of the REFLEX survey \citep[$z<0.3;$][]{reflex}, and the dashed line shows the results of the BCS survey \citep{1997ApJ...479L.101E}, illustrating the current uncertainty on the shape of the luminosity function at $z<0.3$.}
     \label{xlx}
\end{figure}

In Fig.~\ref{xlx} we compare the X-ray luminosity function (XLF) based on the complete part of the catalog with previous determinations.
We used $z<0.04$ for the high-$L_{\rm X}$ part of the XLF, $z<0.03$ limits for $L_{\rm X}<0.7\times 10^{43}$ erg s$^{-1}$, and $z<0.015$ for $L_{\rm X}<8\times 10^{41}$ erg s$^{-1}$. We only used data with absolute values of Galactic latitude above $10^{\circ}$ (avoiding the Galactic plane, where X-ray sensitivity is lower and the identification is incomplete), which makes the fraction of the sky used equal to 0.41. 
Our XLF estimates agree well with previous XLF results, sitting in between REFLEX \citep[$z<0.3;$][]{reflex} and BCS \citep{1997ApJ...479L.101E} results. We note that the local Universe is located in the under-density known as the Keenan-Barger-Cowie void \citep{2013ApJ...775...62K}, which also results in a lower density of massive systems.

\begin{figure}[hbt!]
\centering
   \includegraphics[width=\columnwidth]{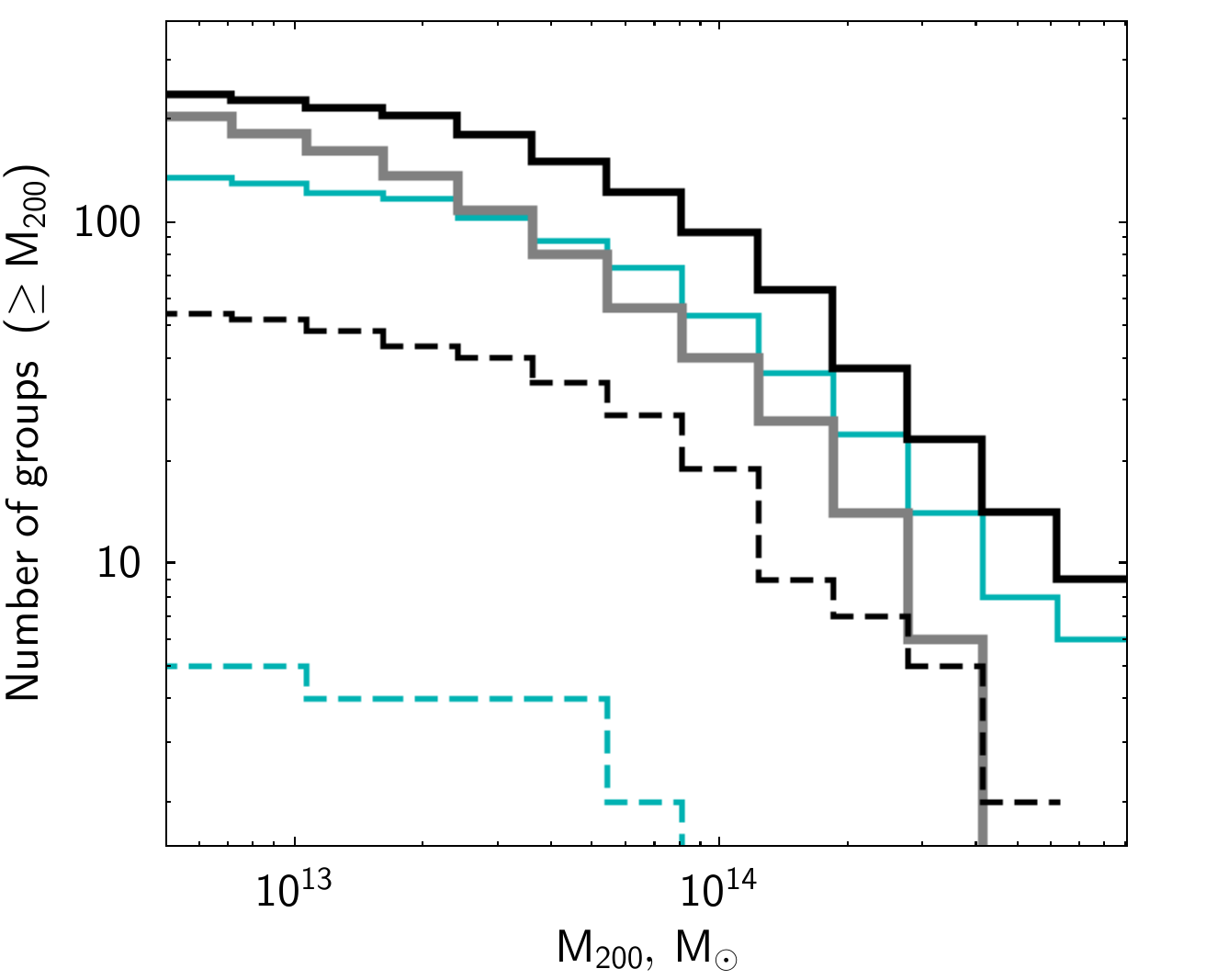}
     \caption{eRASS1 detections (the solid black histogram shows the main catalog and the cyan one shows the identification on both large and small scales) vs. X-ray-unidentified 2MRS groups (solid gray histogram) as a function of their estimated dynamical masses. Dashed black (cyan) histograms show the random identifications for the main (the intersection) catalog. Only 2MRS groups with more than three members are plotted.}
     \label{Fopt}
\end{figure}

\begin{figure}[hbt!]
\centering
   \includegraphics[width=\columnwidth]{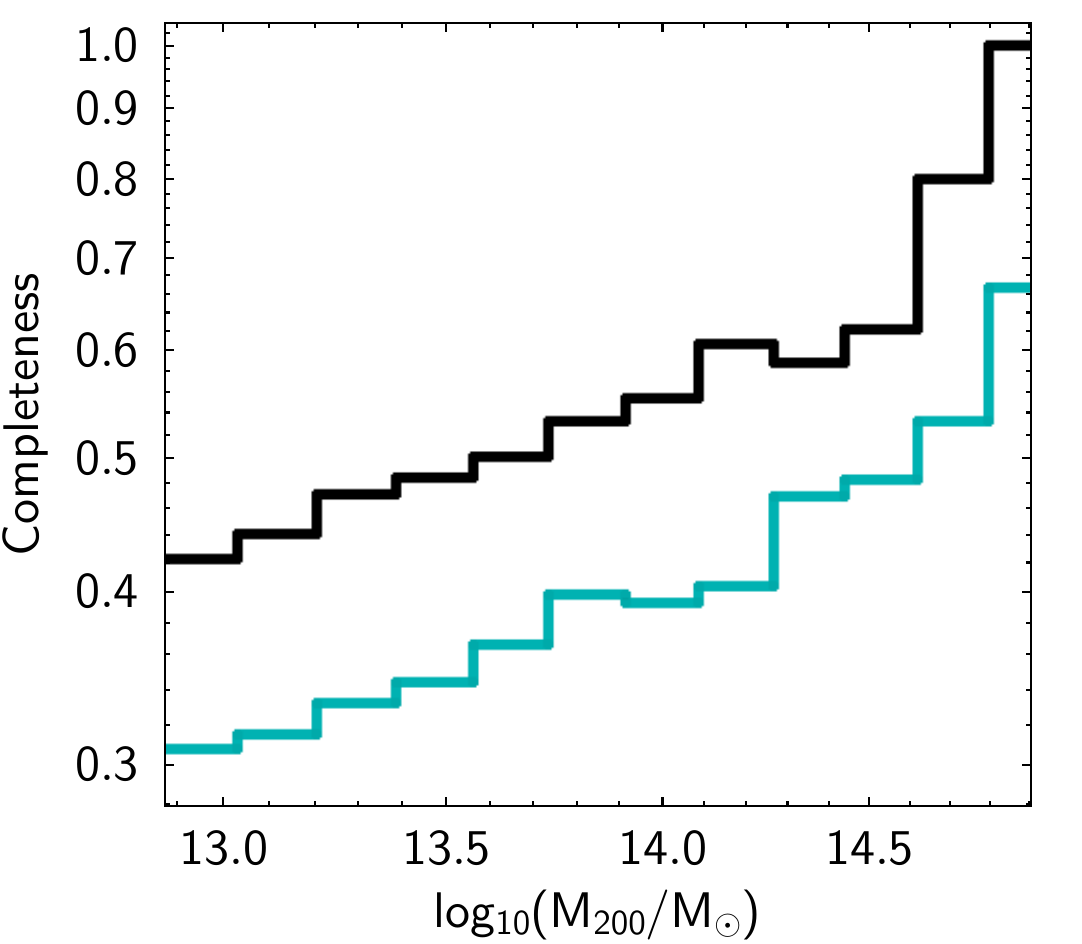}
     \caption{Completeness of eRASS1 detections for the main catalog (black histogram) and the intersection catalog (cyan histogram) as a function of the dynamical mass of the group. Only 2MRS groups with more than three members are plotted.}
     \label{Fcompl}
\end{figure}

\begin{figure}[hbt!]
\centering
   \includegraphics[width=\columnwidth]{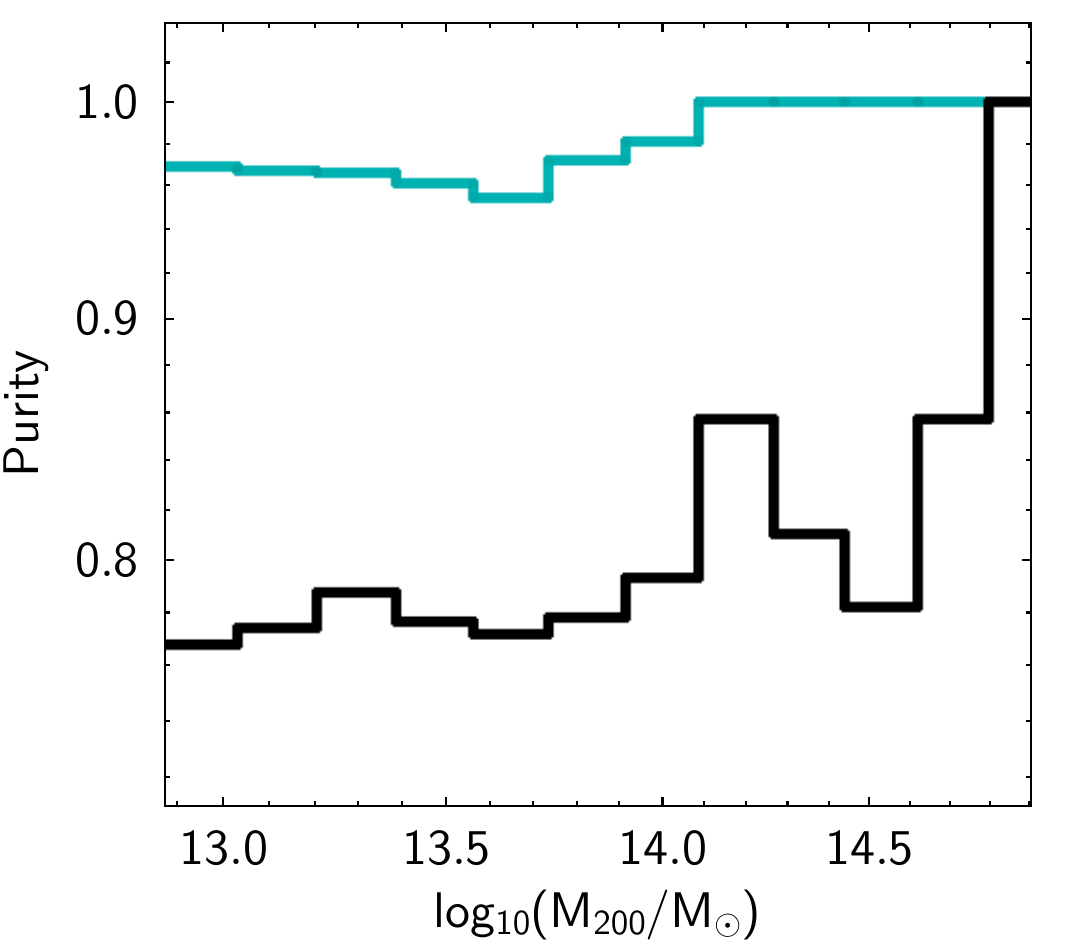}
     \caption{Purity of eRASS1 detections for the main catalog (black histogram) and the intersection catalog (cyan histogram) as a function of the dynamical mass of the group. Only 2MRS groups with more than three members are plotted.}
     \label{Fpure}
\end{figure}

The 2MRS group catalog also provides dynamical mass estimates, which allows us to compute the fraction of identified optical groups as a function of their mass. We chose to limit this comparison to groups with at least four members to reduce the effect of statistical scatter in the mass estimates. Figures~\ref{Fopt}--\ref{Fpure} present the results of this comparison. In Figs.~\ref{Fcompl} and \ref{Fpure} we show  that completeness and purity increase with increasing halo mass. At group masses above $2\times10^{13}~M_\odot$ we provide identification of over 50\% of the groups; even at $1\times10^{13}~M_\odot$ the fraction of identification remains high (42\%).  We find that random identifications are the limiting factor in increasing completeness, as seen from the numbers of undetected and randomly identified groups plotted in Fig.~\ref{Fopt}. {Such a high completeness has also been seen by \cite{clogs}, who performed complete X-ray follow-up of local group sample. Although \cite{2024MNRAS.527..895P} reports a much lower fractional match for GAMA survey, their X-ray  stacking results agree well with the detailed study of this sample \citep{Khalil26}. It has to do with a much higher redshift of the groups in their study, which are well detected with deep \textit{XMM-Newton} exposures \citep{Finoguenov09, 2015A&A...576A.130F}. } 


\begin{figure}[hbt!]
\centering
   \includegraphics[width=\columnwidth]{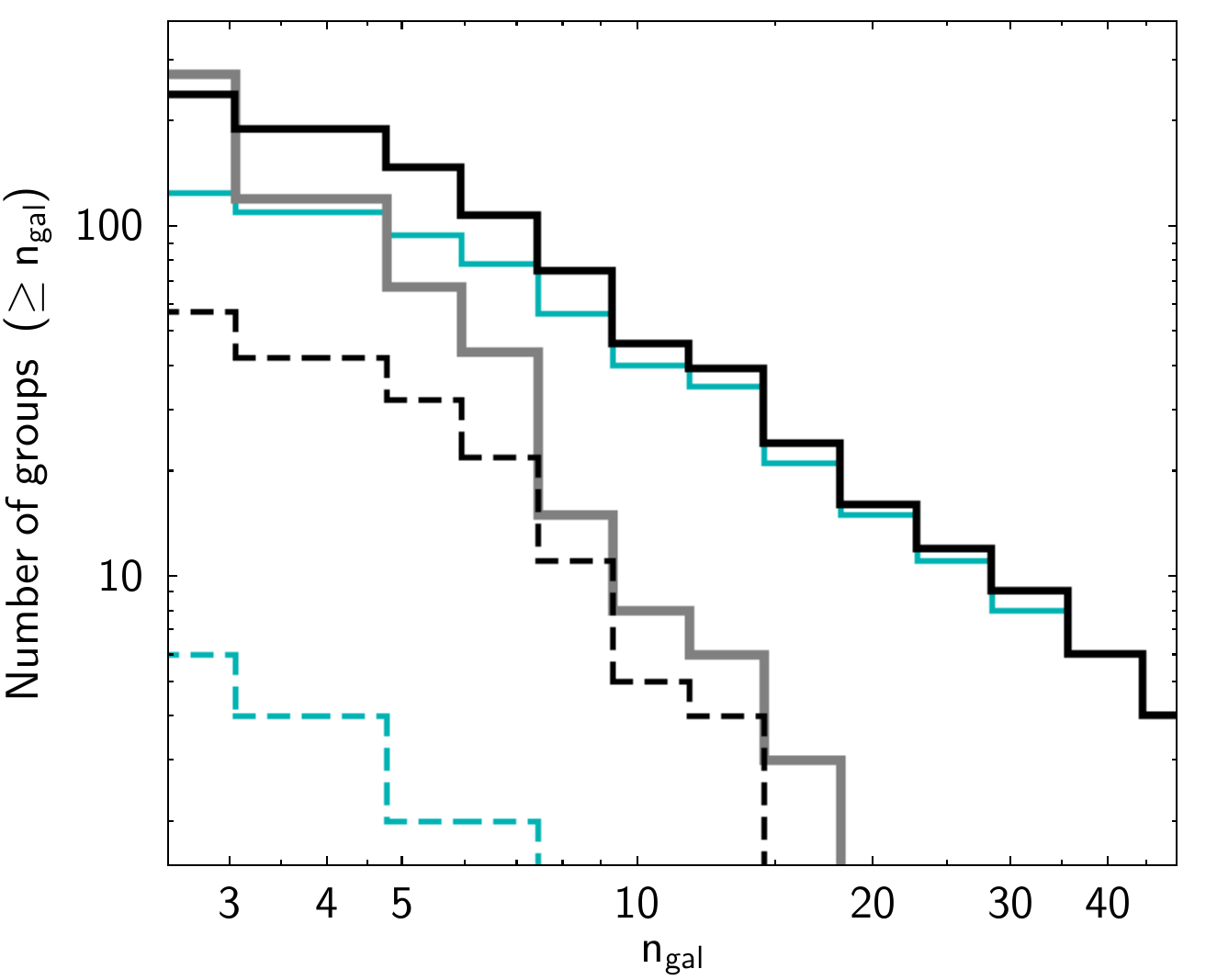}
     \caption{eRASS1-2MRS main catalog (solid black histogram) vs. unidentified 2MRS groups (gray histogram; only objects with dynamical masses above $2\times 10^{13}~M_\odot$ are plotted). The dashed histogram shows the expected random identifications. The comparison demonstrates that the main limitation in identification is purity.}
     \label{Fngal}
\end{figure}

\begin{figure}[hbt!]
\centering
   \includegraphics[width=\columnwidth]{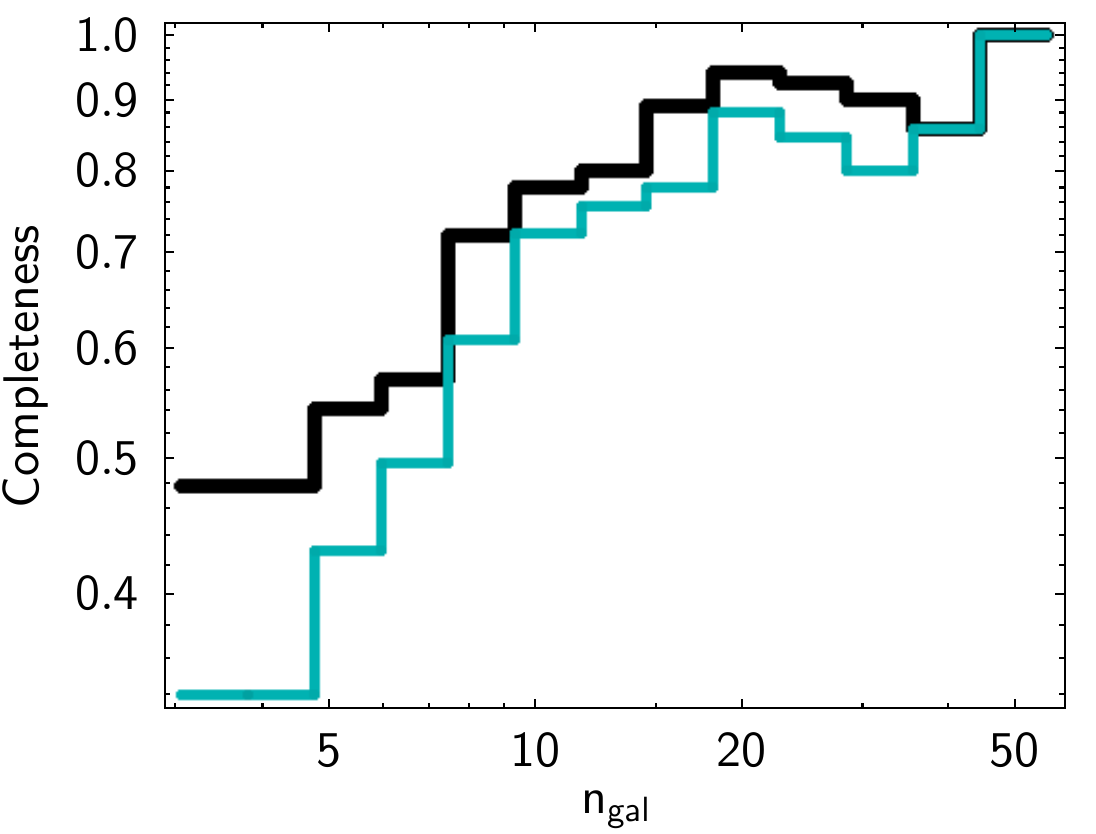}
     \caption{Completeness of eRASS1 detections for the main catalog (black histogram) and the intersection catalog (cyan histogram) as a function of the number of group members. Only 2MRS groups with dynamical mass above $2\times10^{13}~M_\odot$ are plotted.}
     \label{Cngal}
\end{figure}

\begin{figure}[hbt!]
\centering
   \includegraphics[width=\columnwidth]{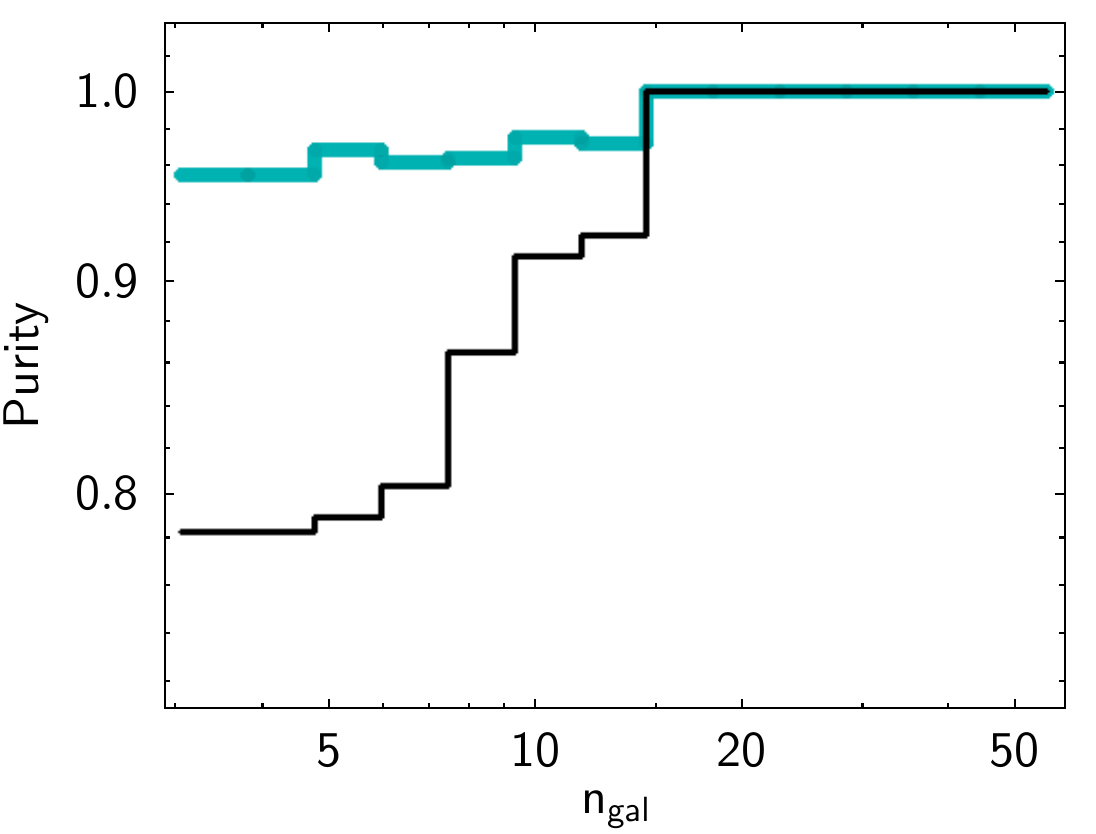}
     \caption{Purity of eRASS1 detections for the main catalog (black histogram) and the intersection catalog (cyan histogram)  as a function of the number of group members. Only 2MRS groups with dynamical masses above $2\times10^{13}~M_\odot$ are plotted.}
     \label{Pngal}
\end{figure}

If we concentrate on massive groups and clusters, having a dynamical mass estimate $>2\times 10^{13}~M_\odot$, we see in Figs.~\ref{Fngal}--\ref{Pngal} that with a larger number of galaxies, our completeness and purity improve. With ten member galaxies, the completeness of our catalogs reaches 80\% and purity 92\% for the full catalog, while the intersection catalog has high purity levels also at low numbers of member galaxies. High values of completeness (at the level of 50\%) for X-ray groups have previously been reported by \cite{Erfanianfar2019} in the AEGIS field, which is an example of a deep survey. 

\section{Sensitivity of the 2MRS group catalog}\label{s:modeling}

\begin{figure}[]
\centering
   \includegraphics[width=\columnwidth]{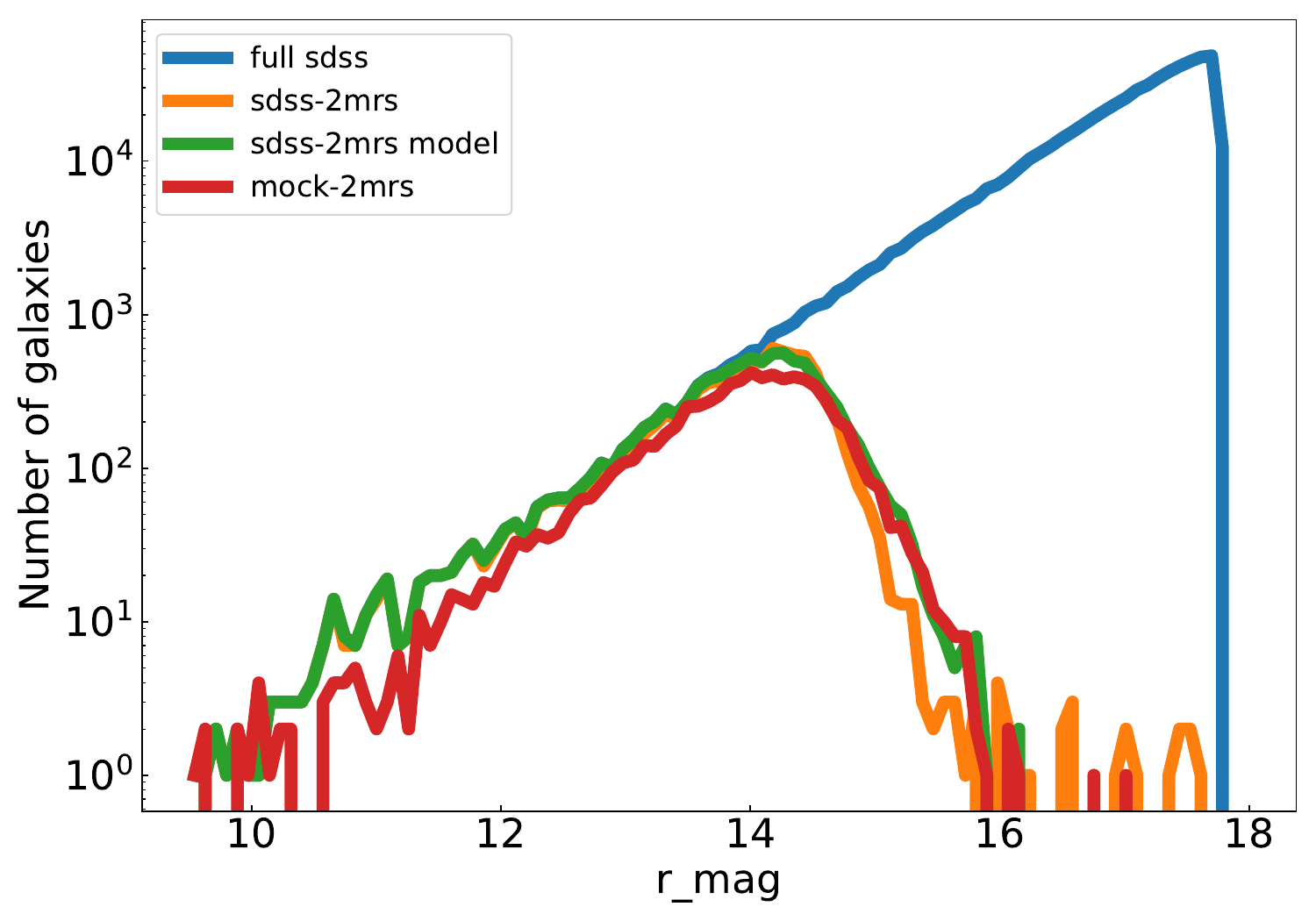}
     \caption{Number of galaxies in the real SDSS (blue), matches from 2MRS (orange), the mock 2MRS (red), {and the completeness model in Eq. \ref{eq:compl_rband} applied to the simulations (green)} as a function of $r$-band magnitude.}
     \label{fig:mock2mrs_gal}
\end{figure}

We studied the sensitivity of the 2MRS group catalog with simulations. We used the \texttt{UchuuSDSS} mocks from \cite{DongPaez2024MNRAS_UchuuSDSS}, a collection of catalogs constructed with an abundance matching scheme to replicate the properties of SDSS. They were used by \cite{Seppi2025A&A...699A.206S} to model the selection function of AXES-SDSS. Since 2MRS is shallower than SDSS, we downgraded the SDSS mock to obtain a 2MRS mock as follows. We started from the real 2MRS galaxy catalog and look for matches in the SDSS catalog within $1^{\prime\prime}$. This gave us the fraction of matched sources as a function of $r$-band magnitude. We modeled this completeness with a sigmoid function:
\begin{equation}
    C = \frac{1}{1 + \exp[\alpha (m_r - m_0)]},
    \label{eq:compl_rband}
\end{equation}
where $\alpha = 5.287\pm0.001$ and $m_0 = 14.409\pm0.001$.
We used this model to sub-select a 2MRS mock from \texttt{UchuuSDSS} by applying rejection sampling: for each galaxy in \texttt{UchuuSDSS}, we drew a random number between 0 and 1. If this number is lower than the probability of being included in 2MRS computed from Eq.~\ref{eq:compl_rband}, we kept the galaxy; otherwise, we discarded it. This method provides a sample with the same properties as the real 2MRS by construction. The result is shown in Fig.~\ref{fig:mock2mrs_gal}.

\begin{figure}[]
\centering
   \includegraphics[width=\columnwidth]{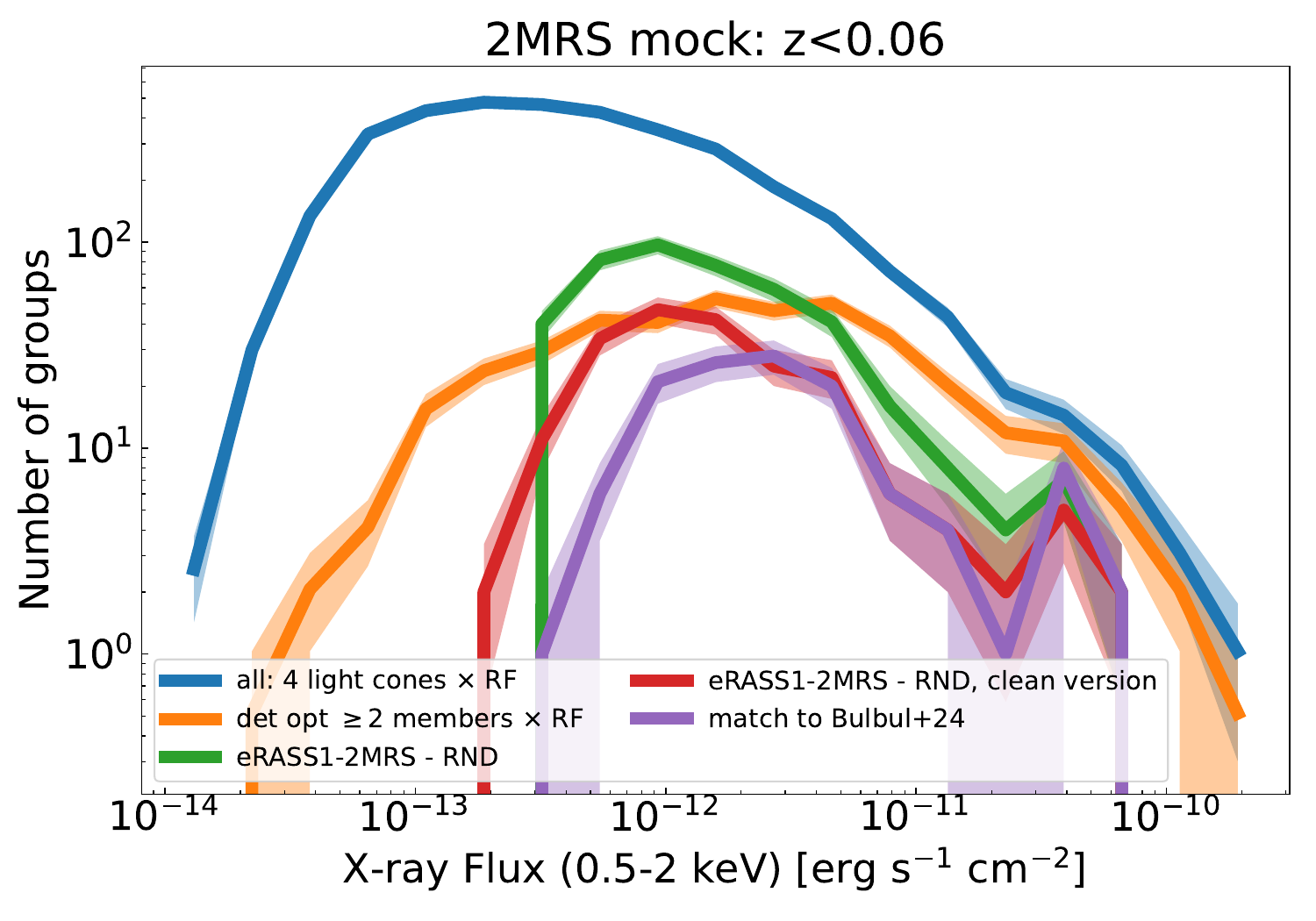}
   \includegraphics[width=\columnwidth]{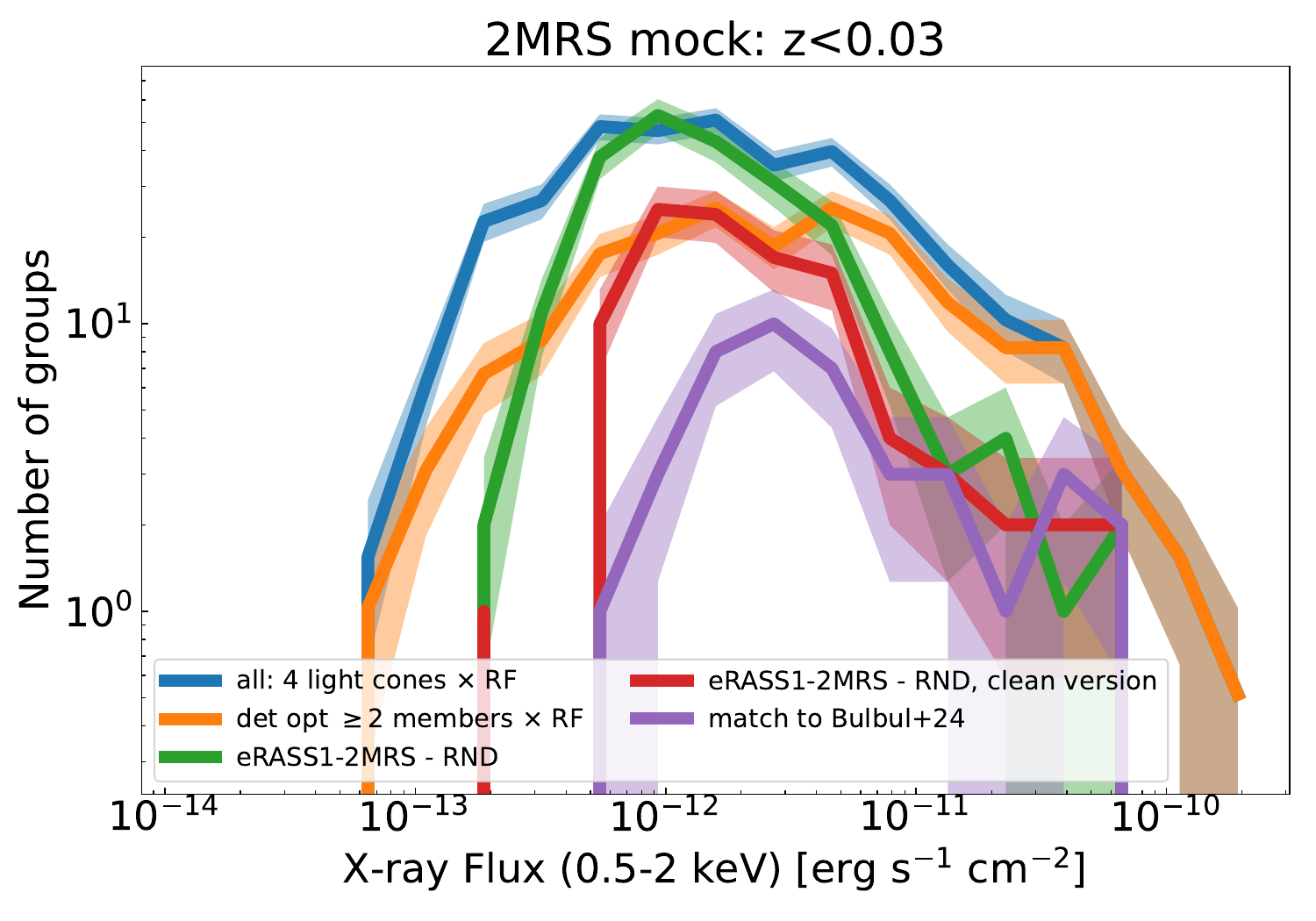}
     \caption{Number of detected groups as a function of flux in the 0.5--2.0~keV band. \textit{Top}: $z<0.06$ (nearly the full sample). \textit{Bottom}: $z<0.03$. The X-ray mock is shown in blue, and a subsample of the optical 2MRS mock groups with at least two members is shown in orange. The results of this paper are shown in green (full catalog) and red (clean catalog), while the published eRASS1 cluster catalog of \citet{erass1_new} is shown in violet. We see that while at $z<0.03$ we detect nearly all simulated groups, at higher redshifts the detection is driven by the incompleteness of the 2MRS group catalog. In plotting our catalogs we statistically corrected for random identifications by presenting the difference between the real and random catalogs, {so we added -RND to the label to specify that}.}
     \label{fig:mock2mrs_det}
\end{figure}

Using the optical finder from \cite{Tempel2017}, we studied the fraction of real groups with an associated detection in the mock 2MRS sample. We also used the ROSAT mocks from \cite{Seppi2025A&A...699A.206S} to simultaneously assign X-ray properties to the groups. We combined the four independent light cones generated in \cite{Seppi2025A&A...699A.206S} to increase the number of groups used to estimate the detection probability. This compensates for the low number of halos in a single realization due to the small cosmological volume probed by 2MRS. We converted fluxes from the $0.6$--$2.3$ keV band to the $0.5$--$2.0$ keV band using XSPEC, assuming an absorbed thermal plasma model (\texttt{tbabs*apec}) with fixed temperature, redshift, metallicity, and galactic hydrogen column density. The model normalization is irrelevant, as only flux ratios are used. Fluxes are computed in the two bands with the \texttt{calcFlux} routine, and the conversion factor is defined as $C = F_{0.5\text{--}2.0} / F_{0.6\text{--}2.3}$. This factor is then applied to rescale the measured fluxes between the two energy bands.
In Fig.~\ref{fig:mock2mrs_det} we illustrate two main points. If we take the nearly full catalog ($z<0.06$ plot), we see that X-ray detections are driven by the sensitivity of the 2MRS two-member groups catalog. At $z<0.03$, the detection of X-ray groups matches all simulated groups, as the 2MRS sensitivity toward detection of two-member groups does not introduce a strong effect. While the data and simulations do not provide an exact match, they capture the trends with the data well. The variance seen for bright sources can be assigned to the low number of clusters in the local volume compared to simulations. {We further illustrate the point that only massive systems are identified at $z>0.04$ by considering the statistical distribution of the detections. At $z>0.04$ we have fewer low-significance sources compared to $0.01<z<0.03$ subsample.}

\begin{figure}[hbt!]
\centering
   \includegraphics[width=\columnwidth]{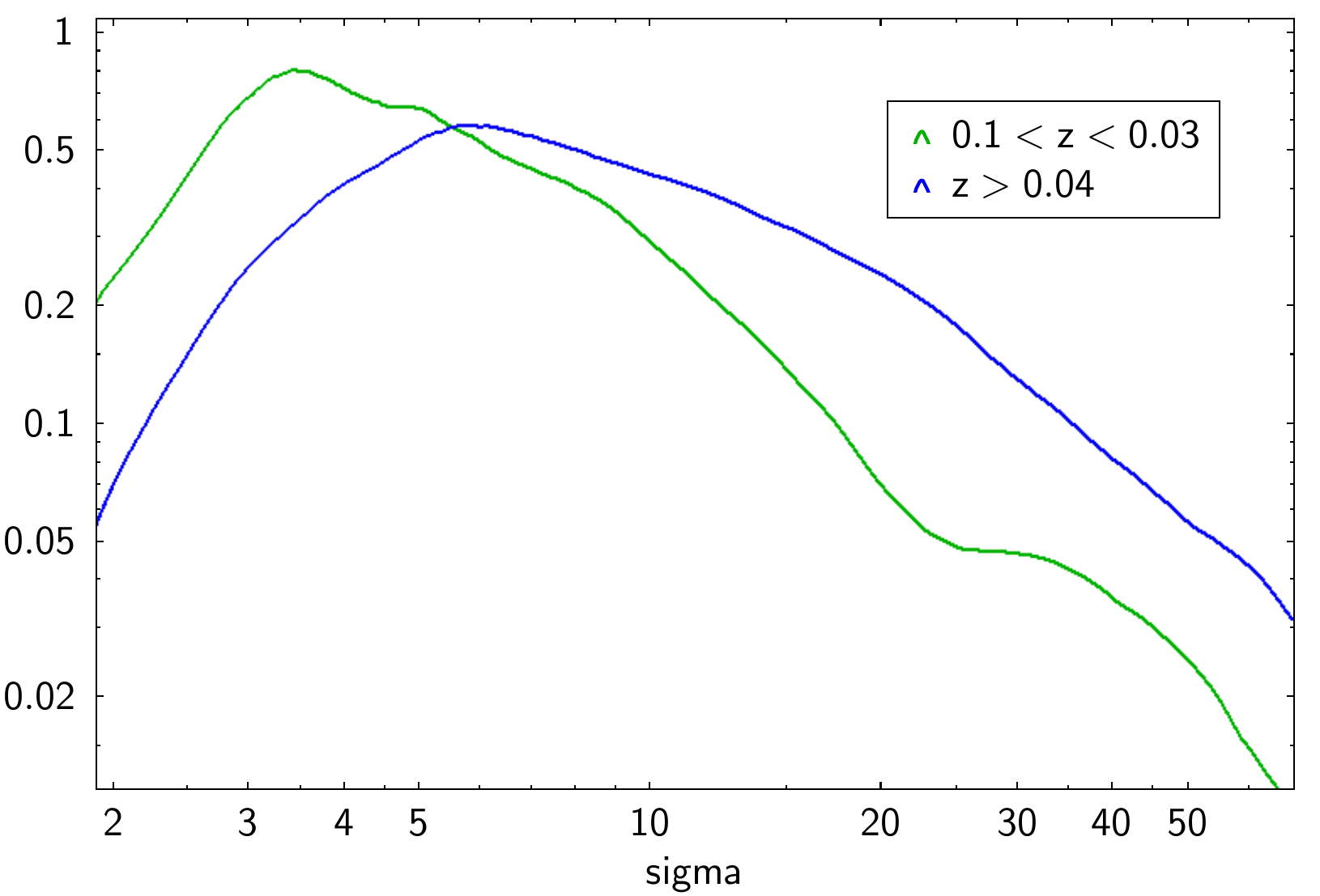}
     \caption{Significance of the flux estimate for eRASS1-2MRS sample. The green curve shows the $0.01<z<0.03$ range, and the blue curve shows the $z>0.04$ range. }
     \label{sigma}
\end{figure}

We used a $5^\prime$ radius and redshift match to find 160 eRASS1 sources in the catalog of \cite{erass1_new} at the same redshift embedded in our sources. Some of them report a factor of 10 less flux, which makes them an identification of a subcomponent, so we cleaned them to be within a factor of 3 in flux, leaving 138 sources. Direct matching to 2MRS groups using only optical group centers yields 122 matches, using similar criteria on angular separation and a redshift. The resulting flux distribution of sources is shown in Fig.~\ref{fig:mock2mrs_det}.  We note, however, that redMaPPer identification of our sources leads to a very similar catalog compared to the official one \citep{2026arXiv260702706D}, indicating that the limitations of the official catalog might also stem from cluster identification. Comparison to the official catalog allows us to conclude that the lack of bright sources is not unique to our study.

\section{Conclusions and summary}
\label{conclusion}

We have presented a new catalog of eRASS1-2MRS X-ray galaxy groups selected based on the baryonic content at $R_{500}$, and we have examined their properties. We have significantly enhanced the representation of the under-explored low-redshift, low-luminosity galaxy groups. This work substantially expands upon previous catalogs available from RASS. We show that with the inclusion of two-member groups we are able to recover the typical XLF of groups, and we confirm this via detailed modeling of the properties of 2MRS galaxy groups. Using the 2MRS group catalog has allowed us to explore the full depths of eRASS1 data and present a catalog of nearby X-ray groups down to lower fluxes than otherwise available \citep{erass1_new}.

Using dynamical mass estimates of groups with at least five members, we show that eRASS1 provides X-ray counterparts to a dominant fraction of galaxy groups, extending down to a mass of $10^{13}~M_\odot$.

\section*{Data availability}

The catalogs of X-ray properties of the eRASS1-2MRS galaxy groups are available at the CDS via anonymous ftp to \url{cdsarc.cds.unistra.fr}.
\begin{acknowledgements}
{We thank the referee for their insightful, detailed suggestions and for the time and effort invested in refining our manuscript.} 
This work is partially based on data from eROSITA, the soft X-ray instrument aboard SRG, a joint Russian-German science mission supported by the Russian Space Agency (Roskosmos), in the interests of the Russian Academy of Sciences represented by its Space Research Institute (IKI), and the Deutsches Zentrum für Luft- und Raumfahrt (DLR). The SRG spacecraft was built by Lavochkin Association (NPOL) and its subcontractors and is operated by NPOL with support from the Max Planck Institute for Extraterrestrial Physics (MPE). The development and construction of the eROSITA X-ray instrument was led by MPE, with contributions from their colleagues at Dr. Karl Remeis Observatory Bamberg \& ECAP (FAU Erlangen-Nuernberg), the University of Hamburg Observatory, the Leibniz Institute for Astrophysics Potsdam (AIP), and the Institute for Astronomy and Astrophysics of the University of Tübingen, with the support of DLR and the Max Planck Society.
ET acknowledges funding from the HTM (grant TK202), ETAg (grant PRG3034) and the EU Horizon Europe (EXCOSM, grant No. 101159513). 
\end{acknowledgements}

\bibliographystyle{aa} % style aa.bst
\bibliography{ref} % your references Yourfile.bib
\begin{appendix}
\section{Reevaluation of the background near bright sources}
\begin{figure}[]
\centering
   \includegraphics[width=\columnwidth]{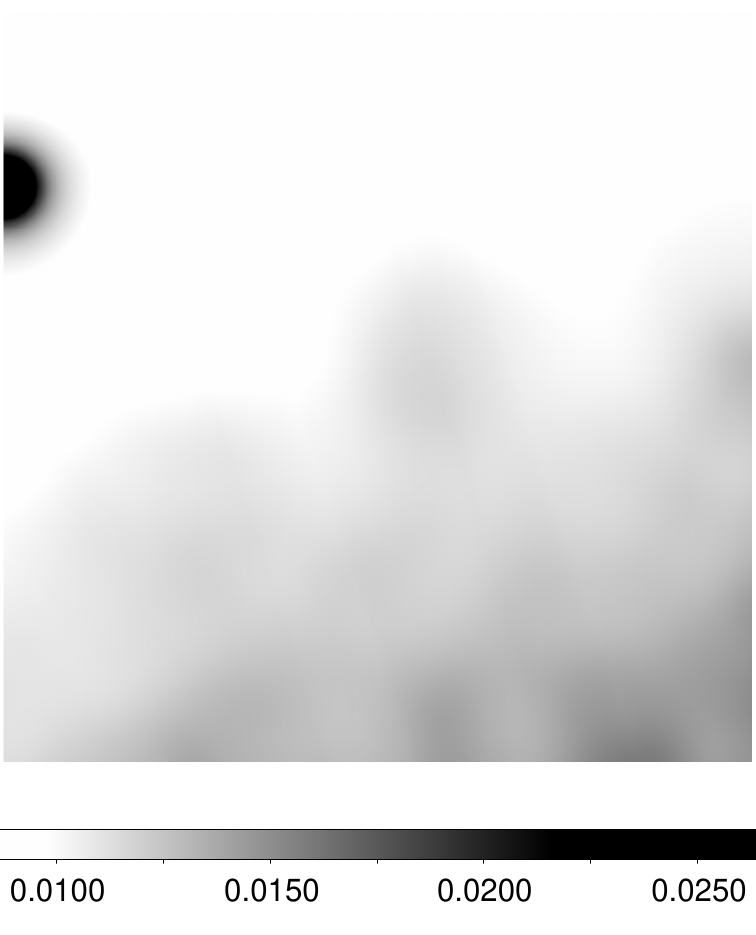}
   \includegraphics[width=\columnwidth]{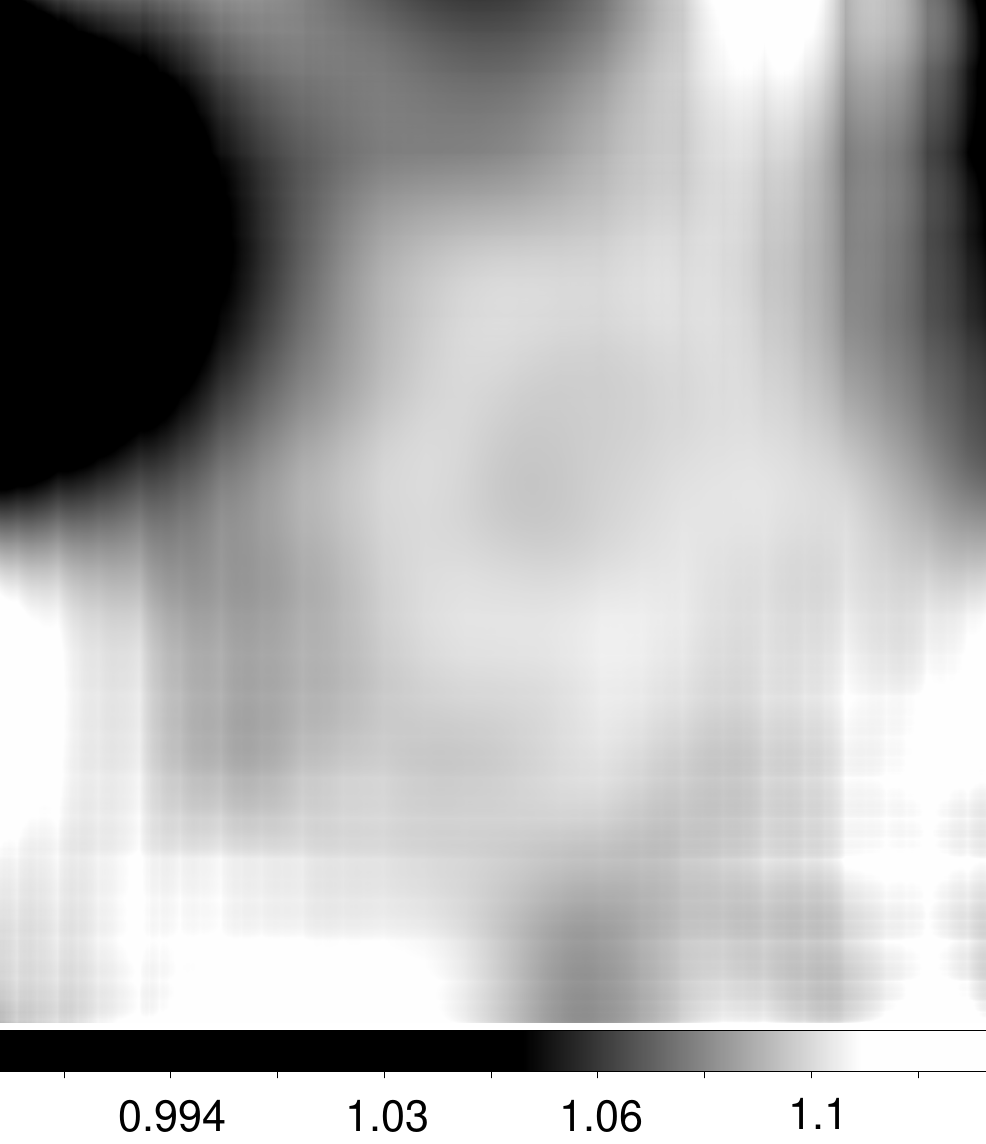}
     \caption{Example of a background problem in tile 150-094. \textit{Top}: Publicly released background. The strongest feature is the source located on the edge of the field (shown in black). \textit{Bottom}: Ratio of the background, reducing the background around the source (shown in black). }
     \label{bgprob}
\end{figure}

The background maps released as a part of eROSITA-DE DR1 were constructed by smoothing of the emission after removal of the bright sources. However, in the case of very bright sources, the wings of eROSITA PSF are too bright and smoothing them overestimates the background, creating large areas of negative net flux, which creates a problem to any analysis. In the wavelet maps it generates artifacts on the largest scales. To address this, we implemented an iterative process to reevaluate the background through wavelet characterization of the bright source emission, and perform multiple iterations. This method provided a more reliable background estimate for our analysis. In total, 16 sky tiles in band 022 were reprocessed using this technique: 093-083, 093-086, 093-194, 093-197,
096-197,
120-108,
132-038,
144-042,
147-140,
150-020,
150-071,
150-094,
153-092,
156-087,
162-122,
and 162-149. We illustrate the background and its change in Fig.~\ref{bgprob}.

\end{appendix}

\end{document}